\documentclass[a4paper,fleqn,usenatbib]{mnras}

\usepackage{newtxtext,newtxmath}

\usepackage[T1]{fontenc}
\usepackage{ae,aecompl}

\usepackage{graphicx}	
\usepackage{amsmath}	
\usepackage{amssymb}	

\usepackage{hyperref}
\usepackage[all]{hypcap}

\title[Baryon effects on void statistics]{Baryon effects on void statistics in the EAGLE simulation}

\author[E. Paillas et al.]{
Enrique Paillas,$^{1, 2}$\thanks{E-mail: epaillas@astro.puc.cl}
Claudia D.P. Lagos,$^{3, 4}$
Nelson Padilla,$^{1, 2}$
Patricia Tissera,$^{5}$ \newauthor
~John Helly$^{6}$
and Matthieu Schaller$^{6}$
\\
\\
$^{1}$Instituto de Astrof\'isica, Pontificia Universidad Cat\'olica de Chile, Av. Vicu\~na Mackenna 4860, Santiago, Chile.\\
$^{2}$Centro de Astro-Ingenier\'ia, Pontificia Universidad Cat\'olica de Chile, Av. Vicu\~na Mackenna 4860, Santiago, Chile.\\
$^{3}$International Centre for Radio Astronomy Research, 7 Fairway, Crawley, 6009, Perth, WA, Australia\\
$^{4}$Australian Research Council Centre of Excellence for All-sky Astrophysics (CAASTRO)\\
$^{5}$Departamento de Ciencias F\'isicas, Universidad Andr\'es Bello, Av. Republica 220, Santiago, Chile\\
$^{6}$Institute for Computational Cosmology, Department of Physics, University of Durham, South Road, Durham, DH1 3LE, UK\\
}

\date{Accepted XXX. Received YYY; in original form ZZZ}

\pubyear{2015}

\begin{document}
\label{firstpage}
\pagerange{\pageref{firstpage}--\pageref{lastpage}}
\maketitle

\begin{abstract}
Cosmic voids are promising tools for cosmological tests due to their sensitivity to dark energy, modified gravity and alternative cosmological scenarios. Most previous studies in the literature of void properties use cosmological N-body simulations of dark matter (DM) particles that ignore the potential effect of baryonic physics. Using a spherical underdensity finder, we analyse voids using the mass field and subhalo tracers in the EAGLE simulations, which follow the evolution of galaxies in a $\rm{\Lambda}$ cold dark matter Universe with state-of-the-art subgrid models for baryonic processes in a $(100 \rm{cMpc})^3$ volume. We study the effect of baryons on void statistics by comparing results with DM-only simulations that use the same initial conditions as EAGLE. When identifying voids in the mass field, we find that a DM-only simulation produces 24 per cent more voids than a hydrodynamical one due to the action of galaxy feedback polluting void regions with hot gas, specially for small voids with $r_{\rm{void}} \le 10\ \rm{Mpc}$. We find that the way in which galaxy tracers are selected has a strong impact on the inferred void properties. Voids identified using galaxies selected by their stellar mass are larger and have cuspier density profiles than those identified by galaxies selected by their total mass. Overall, baryons have minimal effects on void statistics, as void properties are well captured by DM-only simulations, but it is important to account for how galaxies populate DM haloes to estimate the observational effect of different cosmological models on the statistics of voids.\

\end{abstract}




\section{Introduction} \label{Section: Introduction}

In the year 1998, two independent groups of astronomers reported evidence for an accelerated expansion of the Universe, using type Ia supernovae as distance indicators \citep{ Riess1998, 1999ApJ...517..565P}. These results strongly suggest the requirement of a non-zero cosmological constant $\Lambda$ in Einstein's field equations. Together with the fact that as much as 85 percent of the matter in the Universe appears to be dark (only interacting through gravity), the so-called Lambda Cold Dark Matter ($\mathrm{\Lambda CDM}$) scenario has remained popular for being one of the simplest cosmological models that provides a reasonably good account of many key properties of the Cosmos, such as the expansion history of the Universe \citep{1999ApJ...517..565P}, the large-scale structure in the distribution of galaxies (e.g. \citealt{2005MNRAS.362..505C, 2005ApJ...633..560E}) and the existence and structure of the cosmic microwave background \citep{2004ApJ...617L..99O, 2013ApJS..208...19H, 2014A&A...571A..16P}.

The $\mathrm{\Lambda CDM}$ model assumes that general relativity is the correct theory of gravity on cosmological scales. However, alternative approaches that modify the standard theory of gravity have been proposed (e.g. \citealt{2000PhLB..485..208D, 2004LRR.....7....7M, Carroll2005}). One such example is $f(R)$, a family of gravity theories that modify Einstein's Theory of General Relativity by replacing the Ricci scalar $R$ in the Einstein-Hilbert action by an algebraic function of it \citep{Carroll2005}.

Since the solar-system and cosmological tests set tight constraints on the feasibility of modifications to gravity (e.g. \citealt{2011IJMPD..20.1357G, 2014IJMPD..2350036G}), $f(R)$  models have incorporated ``chameleon'' mechanisms that have the effect of removing deviations with respect to general relativity in regions where the gravitational potential is deep enough, such as in our Galaxy \citep{2008PhRvD..78j4021B}. Modified gravity models can produce expansion histories very similar to $\mathrm{\Lambda CDM}$, so it becomes necessary to find alternative ways to test whether they can constitute a better match to our Universe.


A promising way to constrain modified gravity models consists in studying regions of low density, where the chameleon mechanism is effectively suppressed. Cosmic voids are the most prominent under-dense regions in our Universe: these are vast volumes with very low galaxy and mass densities, surrounded by the walls and filaments of the large-scale cosmic web. Previous theoretical studies have shown that voids might occupy more than 50 percent of the total volume of the Universe \citep{El-Ad1997, 2002MNRAS.330..399P, 2014MNRAS.441.2923C}, and their low matter density makes them useful tools for cosmological tests due to their sensitivity to dark energy \citep{2011MNRAS.411.2615L, 2012MNRAS.426..440B, Pisani2015, 2016MNRAS.463..512D}, modified gravity \citep{2012MNRAS.421.3481L, 2013MNRAS.431..749C, 2015MNRAS.451.1036C, 2015MNRAS.451.4215Z, 2016PhRvD..94j3524A} and alternative cosmological scenarios \citep{2015JCAP...11..018M, 2016JCAP...11..015B, 2015JCAP...08..028B}.


Most studies of voids in the literature are based on cosmological N-body simulations that follow the gravitational interaction of dark matter. The baryonic component of the Universe has been so far ignored in these works, and although dark matter accounts for a large fraction of the mass of the Universe, baryons play an important role in a cosmological context.

\citet{2015MNRAS.448.2941S} showed that because of cosmic reionization, most haloes below $\rm{3 \times 10^9\ M_{\odot}}$ do not contain observable galaxies. This breaks the assumptions of the commonly used abundance matching method, which relies on the premise that structure formation can be represented by dark matter-only simulations and that every halo hosts a galaxy. \citet{2014MNRAS.442.2641V} found that gas expulsion and the associated dark matter expansion induced by supernova-driven winds are important for haloes with masses $\rm{M_{200} \le 10^{13}\ M_{\odot}}$, lowering their masses by up to 20 per cent relative to a DM-only model. \citet{2015MNRAS.451.1247S} found that the reduction in mass can be as large as 30 per cent of haloes with $\rm{M_{200} \le 10^{11}\ M_{\odot}}$. They also found that baryons can affect the inner density profile of dark matter haloes, leading to cuspier profiles in the centre due to the presence of stars. Feedback mechanisms triggered by baryons can expell gas from galaxies, even polluting voids with processed material, as suggested by \citet{2016MNRAS.457.3024H}. One would intuitively expect that if feedback processes are strong enough, voids would be more polluted with baryons and hence their properties might differ from their dark matter-only counterparts. It might also be the case that cooling alters the spatial distribution of gas around voids. This modification of the mass distribution could have consequences for the weak lensing signal measured around voids.

In this work we study the effects of baryonic physics on the properties of cosmic voids. We search for voids using the mass field and subhalo tracers in the Evolution and Assembly of Galaxies and their Environment (EAGLE) simulations \citep{2015MNRAS.446..521S, Crain2015}, a suite of cosmological, hydrodynamical N-body simulations that follow the evolution of baryonic and dark matter particles in a $\Lambda \textrm{CDM}$ universe. EAGLE consists of multiple simulations that were run with different mass resolutions, volumes and physical models (the largest simulation consisting on a 100 Mpc on a side comoving box), being one of the first projects that allows the study of baryonic processes in a large simulated volume. The hydrodynamical simulations in the suite have counterparts that were run with the same initial conditions but only following the evolution of dark matter. By comparing these different models we can study the role of baryons in the context of the cosmic web.

EAGLE presents a unique opportunity to explore the effect of baryons on void regions for various reasons. The simulations implement state-of-the-art subgrid models that follow star formation and feedback processes from stars and active galactic nuclei (AGN). These subgrid models, together with a high mass resolution, allow us to trace the distribution of gas and dark matter at different scales in detail, which is a key point in our study given the discussion above.

Another advantage of the EAGLE simulations is that they reproduce the present day stellar mass function, galaxy sizes, and many other properties of galaxies and the intergalactic medium with very good precision (e.g. \citealt{2015MNRAS.446..521S, 2015MNRAS.450.4486F, 2016MNRAS.456.1235S, 2016MNRAS.456.1115B, 2016MNRAS.tmp..895T}). This good agreement with observations becomes important when trying to extrapolate results inferred from voids identified in these simulations to the real Universe. With EAGLE, we can also mimic some of the selections of galaxies in the observations that are used to find voids. In fact, we will show that voids found using galaxies selected by their stellar mass are different than those found if galaxies are selected by their total mass.

We identify voids using a modified version of the algorithm presented in \citet{2005MNRAS.363..977P} (from hereon mP05), which searches for spherical under-dense regions in simulations, either using the mass field or halo tracers. Many void finders have been presented in the literature, which use different tracers to define voids (e.g. \citealt{2002MNRAS.330..399P, 2002ApJ...566..641H, 2002MNRAS.332..205A, 2005MNRAS.360..216C, 2007MNRAS.375..184B, 2007MNRAS.380..551P, 2008MNRAS.386.2101N}). Watershed based methods, such as the Watershed Void Finder \citep{2007MNRAS.380..551P} and ZOBOV \citep{2008MNRAS.386.2101N} are suitable when studying the spatial structure of the cosmic web in detail, as they are parameter free and they do not make assumptions about void shapes or topology. Voids identified with these finders usually exhibit a smooth transition from the under-dense region of a void to the average density. Other algorithms, such as the one we employ in this work, search for spherical regions that satisfy some density criteria. These void regions are usually characterised by a fast transition to the average density, and as a consequence, they are suitable for weak lensing studies (e.g. \citealt{2017MNRAS.465..746S}). In particular, mP05 has been shown to effectively capture differences between $\rm{\Lambda CDM}$ and $f(R)$ models using void statistics in \citet{2015MNRAS.451.1036C}. A comprehensive comparison of different void finding methods can be found in \cite{2008MNRAS.387..933C}.

It is important to mention that here we adopt a cosmological model ($\rm{\Lambda CDM}$) and study the effects of baryons on that particular model. It is difficult to predict what the impact of baryon effects would be if we adopted a different cosmology, but future works could expand on this matter.

The paper is organised as follows: In Section \ref{Section: Simulation and methods} we describe the EAGLE simulations, the void finding algorithm and our methodology to identify voids in the simulations. In Section \ref{Section: Visual inspection of baryon effects} we visually explore the effects on baryons on the large-scale distribution of matter in the simulations, emphasizing on the effects of galaxy feedback on void regions. We present the main results of void statistics in Section \ref{Section: Void abundance}, \ref{Section: Void profiles}  and \ref{Section: Voids and their large-scale environment}. Conclusions and discussions about the work are presented in Section \ref{Section: Conclusions and discussion}. In the Appendix we show the evolution of some of the results found in previous sections at higher redshift, and we explain the methods employed for the calibration of the void finder and the error estimation in detail.

\section{SIMULATIONS AND METHODS} \label{Section: Simulation and methods}

\subsection{Simulation overview} \label{Subsection: Simulation overview}

\subsubsection{The EAGLE project} \label{Subsubsection: The EAGLE project}


EAGLE \citep{2015MNRAS.446..521S} is a project of the Virgo Consortium\footnote{http://www.virgo.dur.ac.uk}, consisting in a suite of cosmological, hydrodynamic N-body simulations of a flat $\rm{\Lambda CDM}$ universe, with parameters inferred from the Planck data \citep{2014A&A...571A..16P}; $\rm{\Omega_{\Lambda} = 0.693}$, $\rm{\Omega_m  = 0.307}$, $\rm{\Omega_b  = 0.048}$, $\rm{\sigma_8 = 0.8288}$, $\rm{n_s = 0.9611}$ and $\rm{H_0 = 67.77\ km\ s^{-1}\ Mpc^{-1}}$. The main Eagle simulation, referred to as L0100N1504, consists of a $\rm{(100\ cMpc)^3}$ volume, initially containing $1504^3$ gas particles with an initial mass of $\rm{1.81\times 10^6\ M_{\odot}}$ and the same number of dark matter particles, with a resolution of $\rm{9.70\times 10^6\ M_{\odot}}$.

Initial conditions were generated at z = 127, using $\rm{2^{nd}}$ order Lagrangian perturbation theory \citep{2010MNRAS.403.1859J}. They were evolved using a parallel N-body smooth particle hydrodynamics (SPH) code, an extensively modified version of Gadget-3, a more computationally efficient version of Gadget-2, described in \citet{Springel2005}. The SPH implementation in EAGLE is referred to as Anarchy (Dalla Vecchia, in preparation. See also \citealt{2015MNRAS.454.2277S}), and it is based on the general formalism described by \citet{2013MNRAS.428.2840H}, with improvements to the kernel functions \citep{2012MNRAS.425.1068D} and viscosity terms \citep{2010MNRAS.408..669C}. Anarchy alleviates the problems associated with standard SPH in modelling contact discontinuities and fluid instabilities.

Subgrid schemes are applied to model astrophysical phenomena below the scales of the resolution of the simulation. These include models for radiative cooling and photo-heating, star formation, stellar mass loss and metal enrichment, stellar feedback from massive stars, black hole growth and feedback from active galactic nuclei (AGN). See \citet{Crain2015} for a description of how the free parameters associated with feedback are calibrated. Each of these subgrid models are summarized as follows:

Cooling and photo-heating is implemented following \citet{2009MNRAS.393...99W}. Under the assumption of ionization equilibrium, exposure to the cosmic microwave background and the ultraviolet and X-ray backgrounds from galaxies and quasars \citep{2001cghr.confE..64H}, the abundance of 11 elements that dominate the cooling rates are tracked, tabulated using Cloudy \citep{1998PASP..110..761F}.

Gas particles are stochastically converted into stars following \citet{2008MNRAS.383.1210S}, using the metallicity-dependent density threshold of \cite{2004ApJ...609..667S}. The star formation rate per unit mass of these particles is calculated with the gas pressure using an analytical formula that reproduces the observed Kennicutt-Schmidt law \citep{1998ApJ...498..541K} in disc galaxies \citep{2008MNRAS.383.1210S}, where the lowest possible gas pressure is set by a polytropic equation of state, $P \propto \rho ^{4/3}$. A \citet{2003PASP..115..763C} initial mass function (IMF) in the range $0.1-100\ \rm{M_{\odot}}$ is adopted, where each particle represents a single age stellar population.

It is assumed that stars with a initial mass above $6\ \rm{M_{\odot}}$ explode as supernovae after $3 \times 10^7\ \rm{yr}$, transferring the energy from the explosions as heat to the surrounding gas. A fraction of the surrounding gas gets an instant raise in temperature of $10^{7.5}\ \rm{K}$. This thermal energy is stochastically distributed among gas particles neighboring the explosion event, without any preferential direction \citep{2012MNRAS.426..140D}. The probability of energy injection depends on the metallicity and density of the local environment in which the star particle formed. The simulated stellar populations release metals to their surrounding environment via type Ia supernovae, winds, supernovae from massive stars, and AGB stars, following the methodology discussed in \citet{2009MNRAS.399..574W}.

The gas particles at the centre of the potential well of haloes that reach masses above $10^{10}\ \rm{h^{-1}M_{\odot}}$ are converted into black hole particles of $10^{5}\ \rm{h^{-1}M_{\odot}}$ \citep{2005MNRAS.361..776S}. These black holes can accrete mass based on the modified Bondi-Hoyle model of \citet{2016MNRAS.462..190R}, as well as undergo merging with other black holes \citep{2009MNRAS.398...53B}. A fraction of 0.015 of the rest-mass energy from the accreted mass is returned to the surrounding medium as energy. AGN feedback is implemented thermally, similar to stellar evolution feedback. Gas particles receive AGN feedback energy stochastically, having their temperatures raised instantly by $10^{8.5}\ \rm{K}$.

In this manuscript we will also use the distribution of subhaloes identified in the simulation to define voids. These subhaloes correspond to gravitationally bound sub-structures that are located within a major dark matter halo. Virialized haloes are identified by applying the friends-of-friends method. Sub-structures within haloes are identified using SUBFIND \citep{2001MNRAS.328..726S, 2009MNRAS.399..497D}. We refer to these sub-structures as subhaloes.

Our results are only valid for voids found using this subhalo finder. Other finders include RockStar \citep{2013ApJ...762..109B}, AHF \citep{2004MNRAS.351..399G, 2009ApJS..182..608K} BDM \citep{1997astro.ph.12217K} and others, but it is beyond the scope of this work to study the impact of the subhalo finding algorithm on the statistics of voids.

\subsubsection{Simulations used in this work} \label{Subsubsection: Simulations used in this work}

As mentioned in the previous section, EAGLE is a suite that consists of simulations run with different resolutions, volumes and physical models.
Here we list the simulations that are used in this work. Note that only the first two simulations listed here are used for constructing void catalogs, while the later are used for visualization purposes.

Simulations used for constructing void catalogs:\

\begin{itemize}

\item Ref-L0100N1504: The main EAGLE simulation. It has a $\rm{(100\ cMpc)^3}$ volume, initially containing $1504^3$ gas particles with an initial mass of $\rm{1.81\times 10^6\ M_{\odot}}$ and the same number of dark matter particles, with a mass of $\rm{9.70\times 10^6\ M_{\odot}}$. The physical model this simulation was run with is referred to as the "reference" EAGLE model, and other simulations run with the same subgrid physics and numerical recipes also include the prefix "Ref" next to its name.

\item DM-L0100N1504: A simulation run with the same initial conditions as Ref-L0100N1504, but that only includes dark matter particles. It contains $1504^3$ dark matter particles with a mass of $\rm{1.44\times 10^6\ M_{\odot}}$.\\

Simulations used for visualization and calibration of the void finding algorithm:\\

\item Ref-L0025N0376: A smaller simulation that has a $\rm{(25\ cMpc)^3}$ volume, initially containing $376^3$ gas particles and an equal number of dark matter particles, with a resolution of $\rm{1.81 \times 10^6\ M_{\odot}}$ and $\rm{9.70\times 10^6\ M_{\odot}}$, respectively.

\item DM-L0025N0376: A simulation run with the same initial conditions as Ref-L0025N0376, but that only includes dark matter particles. It has a $\rm{(25\ cMpc)^3}$ volume, initially containing $376^3$ dark matter particles with a resolution of $\rm{1.44\times 10^6\ M_{\odot}}$.

\item NoFeedback-L0025N0376: A simulation run with a physical variation of the reference model that suppresses stellar and AGN feedback.

\end{itemize}

We search for voids on Ref-L0100N1504 and DM-L0100N1504. As these simulations share the same initial conditions, the comparison of their voids provides us with a direct assessment of the effects of baryons.We make use of the smaller simulations Ref-L0025N0376, DM-L0025N0376 and NoFeedback-L0025N0376 to visually explore the simulation volume and calibrate our void finding algorithm. Note that we do not perform an analysis of void statistics on these smaller simulations, since the statistics become restrictively poor as only a handful of voids are identified in their limited volume.

\subsection{Void identification} \label{Subsection: Void identification}

\subsubsection{Void finding algorithm} \label{Subsubsection: Void finding algorithm}

We use an extensively modified version of the void finder presented in \citet{2005MNRAS.363..977P}, with improvements on its computational efficiency, its convergence for different mass resolutions and box sizes, and adapted to run on parallel computers. Voids can be identified using the mass field or halo tracers in the simulation. This modified version of the void finder will be presented in detail in Paillas \& Padilla (in preparation). Here we briefly describe the algorithm:\\

(i) It searches for low-density regions in the simulation by constructing a rectangular grid, and calculating the number of tracers that fall in each cell of the grid. The centre of a cell that is empty of tracers is considered to be a prospective void centre.

(ii) It measures the under-density in spheres of increasing radius about prospective void centres until some integrated density threshold, $\Delta_{\rm{void}}$, is reached. Here we adopt a value of $\Delta_{\rm{void}} = -0.8$. When the mass field is used for identification, $\Delta_{\rm{void}}$ corresponds to a mass density. When haloes or subhaloes are used as tracers, the threshold corresponds to a number density. Only the largest sphere about any one centre is kept, and the void radius is defined as the radius of that under-dense sphere.

(iii) It rejects all spheres whose centres overlap with a larger neighbouring sphere by more than a given percentage of the sum of their radii. Increasing this percentage will also increase the number of voids in the sample, but it could eventually lead to larger covariances in the results. In this work we construct samples in which we allow two adjacent voids to overlap between a 40 - 20 per cent of the sum of their radii.

The value of $\Delta_{\rm{void}} = -0.8$ is commonly used in the literature of void studies, both in observations and theoretical works. It is motivated by a linear theory argument presented in \cite{1992ApJ...388..234B}, who showed that the void regions we observe in the present time probably correspond to regions with mass densities around 20 per cent of the mean density in the Universe. Choosing a lower density threshold would result in a catalog with fewer, more extreme under-dense regions.

On average, void regions in the Universe exhibit spherical symmetry when stacked, a fact that is used as an argument for constraining cosmological models via the Alcock-Paczynski test (e.g. \citealt{2014MNRAS.443.2983S, 2016PhRvL.117i1302H}). However, individual void shapes can be far from spherical. Since with our finder we search for spherical regions that satisfy an integrated density criteria, it is not suitable for studies that require a detailed description of the geometry of individual voids. Nevertheless, the regions it identifies show a strong density contrast and a fast transition to the average density, which boosts the weak lensing signal measured around them. This information has been proven to be useful for testing modified gravity models in N-body simulations as in \citep{2015MNRAS.451.1036C}. Watershed-based void finders have also been used to disentangle gravity models using void statistics as in \citet{2015MNRAS.451.4215Z}, who used the Void Identification and Examination (VIDE) toolkit \citep{2015A&C.....9....1S} to differentiate models of $f(R)$ gravity from cold dark matter cosmology. Bose et al. (in prepation) will compare different void finding techniques in the context of differentiating between $f(R)$ gravity models, in an attempt to study the impact of the different assumptions that each algorithm uses to define void regions.

Step iii) of the algorithm naturally replaces voids of comparable sizes with similar centres by a larger void that occupies the volume of all the smaller voids together. However, as discussed above, individual void shapes in the simulation can deviate significantly from spherical symmetry, and it is therefore necessary to allow a certain degree of overlap between adjacent spheres to sample the low density regions of the cosmic web to a reasonably good extent. This is something that has to be taken into account during the calculation of errors, since two voids with similar centres may represent only one physical void in the simulation, which could artificially lower the statistical errors. In this work we try different degrees of overlap to study the impact of this parameter on the results.

Voids identified by our algorithm cannot be arbitrarily small. Since a sphere is geometrically defined by four points on its surface, the smallest possible void in our sample would need to have at least four subhaloes on its surface to be well defined. Additionally, it would need to satisfy the $\Delta_{\rm{void}} = -0.8$ density criteria. Thus, when finding voids using subhalo tracers, the minimum radius a void in our sample can have is such that the integrated subhalo number density of the void satisfies the density criteria when it has 4 subhaloes on its surface. This constraint can be written as

\begin{align} \label{Equation: Minimum radius - subhaloes}
\frac{4}{\frac{4}{3} \pi r_{\rm{min}}^3} &=  \frac{\rm{N_{sub}}}{\rm{V_{tot}}} \times | \Delta_{\rm{void}}+1 |\ , \\
r_{\rm{min}}^3 &= \frac{3 \rm{V_{tot}}}{| \Delta_{\rm{void}}+1 | \pi \rm{N_{sub}}}\ \ ,
\end{align}

\noindent where $\rm{V_{tot}}$ is the total volume of the simulation box and $\rm{N_{sub}}$ is the number of subhaloes in the simulation.


Subhaloes correspond to over-dense peaks in the matter distribution of the simulation. When using the mass field (i.e. individual particles) to identify voids, we require a larger number of objects to define the minimum void radius, in order to reduce the number of spurious voids found by the algorithm. Here we require 20 particles to define the minimum void radius, so the constraint reads

\begin{align} \label{Equation: Minimum radius - particles}
\frac{20}{\frac{4}{3} \pi r_{\rm{min}}^3} &=  \frac{\rm{N_{part}}}{\rm{V_{tot}}} \times | \Delta_{\rm{void}}+1 | \\
r_{\rm{min}}^3 &= \frac{15\rm{V_{tot}}}{| \Delta_{\rm{void}}+1 | \pi \rm{N_{part}}}\ \ .
\end{align}

It is also important to mention that the grid that is constructed in step i) of the algorithm determines the resolution at which void centres are identified. Increasing the resolution of this grid also increases the probability of finding the optimal centre of a void in the simulation. However, it also increases the number of spurious void centres that arise because of shot noise. This in turn increases the computational time used by the finder, so there is a trade-off between the optimal grid resolution and the computational resources that are available. If the resolution at which a void centre is identified is poor, the void radius is also going to be affected, since the integrated density threshold about that centre will change. Since identifying voids using subhalo tracers is not computationally expensive, we choose a cell size such that 
the error associated to the radius of the smallest void in our sample is smaller than 10 per cent (consequently, errors for larger voids in the catalog will be smaller than this). Identifying voids using the mass field is more computationally expensive (roughly 40000 hours of CPU time), so we cannot arbitrarily increase the resolution of the grid. We instead perform convergence tests to choose this parameter, as described in Appendix \ref{Appendix: Convergence tests}. The minimum void radius for each void sample can be found in Table \ref{Table: Void catalog}.

\subsubsection{Finding voids in EAGLE} \label{Subsubsection: Finding voids in EAGLE}

\begin{figure}
	\includegraphics[width=\columnwidth]{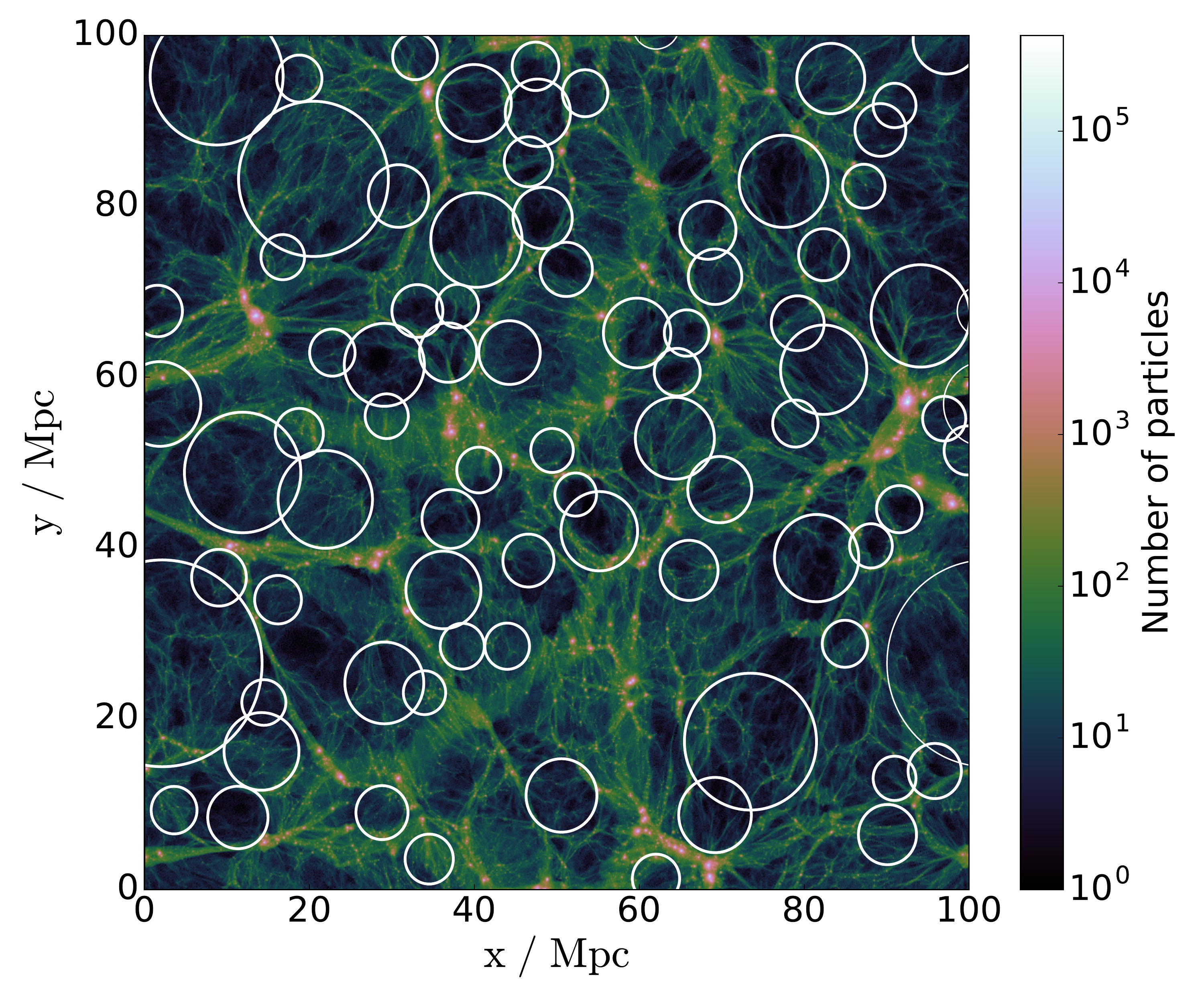}
    \caption{A 1.5 Mpc thick slice along the z-coordinate of DM-L0100N1504 showing the distribution of dark
     matter particles in the simulation. A grid with $(1200)^{2}$ cells on the xy plane was constructed, and the number of particles
     that fall in each cell is shown by the colourmap, using a logarithmic contrast. Voids identified in the mass field
     that have centres within the slice are shown by white circles, where the radius of the circle
     corresponds to the physical radius of the void. We show a void sample with 20 per cent of overlap.
	 Regions that look under-dense but do not feature a circle in the figure can still be embedded in voids that have
     centres outside the slice, or the void associated with them could have been deleted by the overlapping criteria adopted.
     At the same time, circles that appear to surround over-dense regions mainly arise because of projection effects.}
    \label{Figure: Voids in slice}
    \centering
\end{figure}

We run our void finder on Ref-L0100N1504 and DM-L0100N1504 using the mass field (individual particles) and subhalo tracers. When using the mass field, calculating the integrated density profile for each prospective void centre is computationally demanding due to the large number of particles in these simulations. We therefore find the prospective void centres using all the particles in the simulations, but we select a random sub-sample of 1 per cent of the total amount of particles to compute the integrated density profile for each centre. To study the effect of this sub-sampling on the structures that are identified, we calibrate the void finder in Ref-L0025N0376, finding that using this percentage of particles produces relative errors in the radius of the voids of less than ten percent. Considering that L0100N1504 is 64 times bigger in volume than L0025N0376, we expect that voids found in L0100N1504 using 1 per cent of the particles have errors around 1 per cent in the radius compared to those that would be measured using all the particles in the simulation, because of shot noise alone.

When using subhalo tracers, we construct the following subhalo samples to identify voids:

\begin{enumerate}
\item All subhaloes in Ref-L0100N1504 with stellar mass $\ge 10^8\ M_{\odot}$. This results in 40076 subhaloes.

\item The 40076 subhaloes with the largest total mass in Ref-L0100N1504. The total mass of a subhalo is the sum of all the mass contained in gas, dark matter, stellar and black hole particles. This sample was constructed to have the same number density of objects as sample i), but since the scatter between the total and stellar mass can be large \citep{2016MNRAS.461.3457G}, the lowest stellar mass in the sample is $\rm{M_{stel}} = 6 \times 10^7\ \rm{M_{\odot}}$, while the lower total mass is $\rm{M_{stel}} = 1.6 \times 10^10\ \rm{M_{\odot}}$. The stellar mass distribution of this sample peaks at $\rm{M_{stel}} = 2.8 \times 10^9\ \rm{M_{\odot}}$ and tappers off at lower masses.

\item The 40076 subhaloes with most total mass in DM-L0100N1504. Since this simulation does not include baryons, only dark matter contributes to the total mass of a subhalo. The lowest subhalo mass in the sample is $2 \times 10^{10}\ \rm{M_{\odot}}$.

\end{enumerate}

While finding voids using a sample of subhaloes selected by total mass is more related to the approach taken in cosmological simulations, a stellar mass cut is more related to observations. The cut in stellar mass can be compared to a cut in magnitude, as used, for example, by \cite{2013MNRAS.434.1435C} with the Sloan Digital Sky Survey (SDSS) main galaxy sample.

As mentioned in the previous section, we can obtain void samples with different degrees of overlap from our algorithm. To test the effect of the choice of this parameter on the results, we construct void catalogs with 40, 30 and 20 per cent of overlap for each of the samples described above.\\

In Fig. \ref{Figure: Voids in slice} we show a 1.5 Mpc thick slice along the z axis of a snapshot of DM-L0100N1504 at $z=0.0$. Our algorithm successfully identifies under-dense structures across the simulation. Regions that look under-dense but that are not enclosed by a circle could still be identified as voids whose centres do not fall within the ranges of the slice (which is the case for some of the obvious under-dense regions in the figure), or they could have been erased by the overlapping criterion. At the same time, some white circles appear to enclose high density regions, which is mostly caused by a projection effect due to the fact that the radius of the smallest voids are comparable to the thickness of the slice.

\section{Visual inspection of baryon effects} \label{Section: Visual inspection of baryon effects}

\begin{figure*}
	\includegraphics[width=\textwidth]{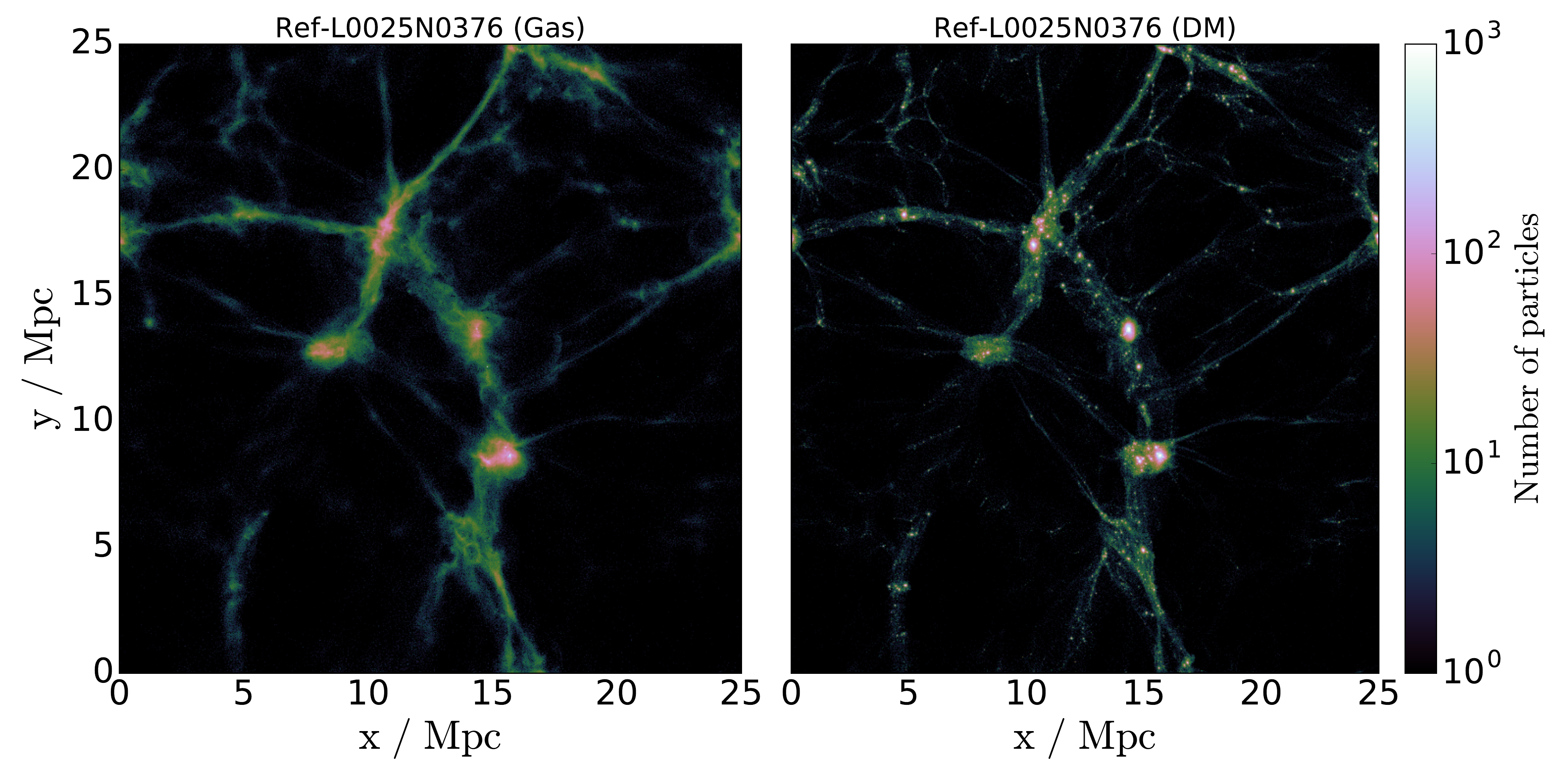}
    \caption{A 2.0 Mpc thick slice along the z-coordinate of Ref-L0025N0376 at z = 0,
    showing the distribution of gas (left panel) and dark matter (right panel)
    particles in the simulation. A grid with $(1200)^{2}$ cells on the xy plane
    was constructed, and the number of particles that fall in each is cell shown
    by the colourmap, using a logarithmic contrast. The distribution of gas appears
    to be more diffuse, with wider filaments and less clumpy regions than the 
    dark matter distribution. This is due to different baryonic processes that take
    place within these regions, such as cooling and photoheating, stellar evolution and
    AGN feedback, all of which affect the properties of gas particles.}
    \label{Figure: Gas vs Dark matter}
    \centering
\end{figure*}

\begin{figure*}
	\includegraphics[width=\textwidth]{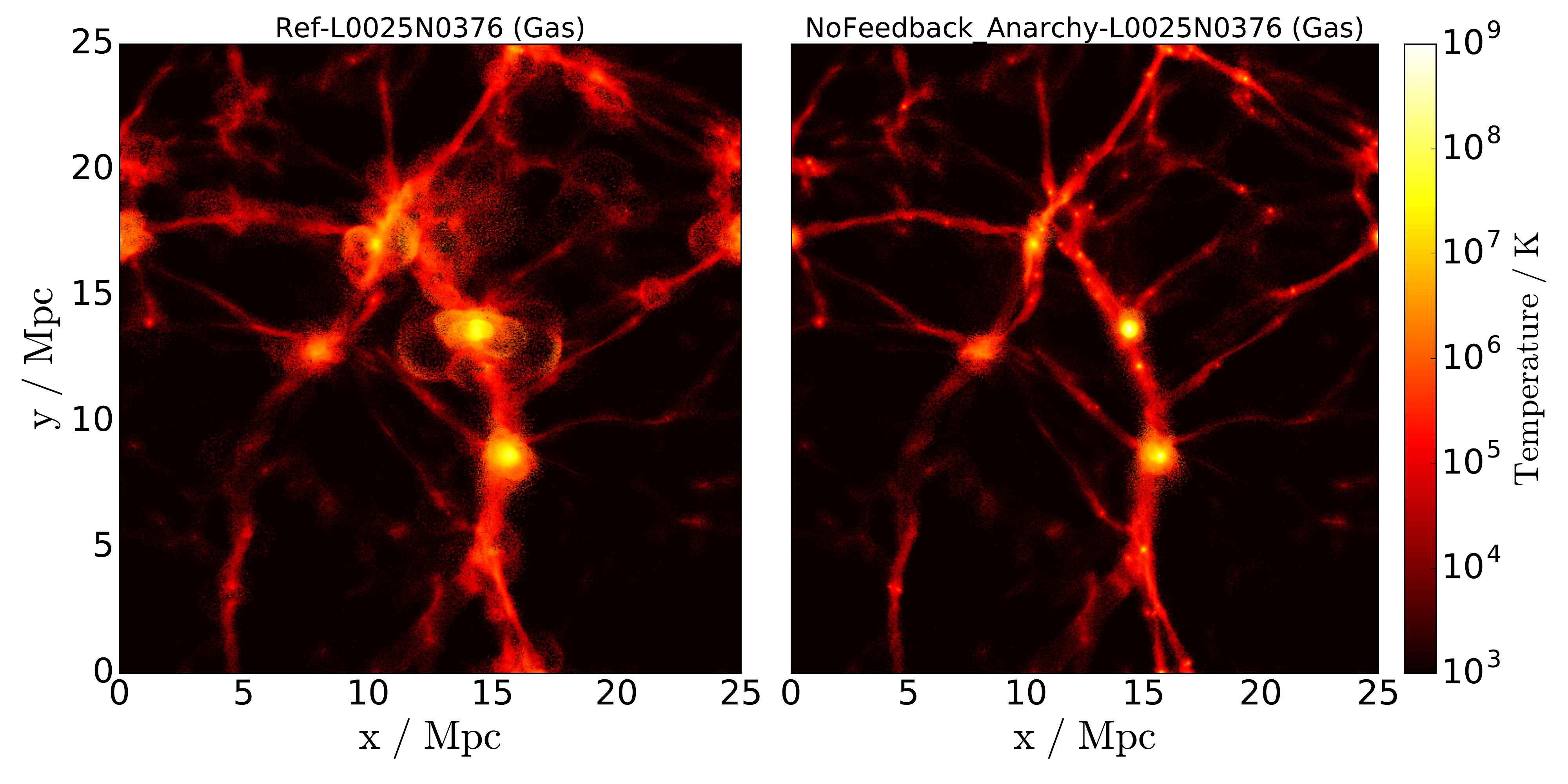}
    \caption{A 2.0 Mpc thick slice along the z-coordinate showing the distribution of gas particles in Ref-L0025N0376 (left panel) and
     NoFeedback\_Anarchy-L0025N0376 (right panel) at z = 0.
     A colourmap with a logarithmic contrast shows the temperature of these gas particles.
     NoFeedback\_Anarchy-L0025N0376 is a physical variation of EAGLE which supresses stellar and AGN feedback in the simulation.
     The simulation with baryonic physics exhibits an over-abundance of particles with high temperatures in regions surrounding
     the dense clumps of matter where haloes are formed. Shock-like structures suggest that part of this mass is being ejected
     from these haloes into the more under-dense regions via feedback events.}
    \label{Figure: Ref vs No Feedback}
    \centering
\end{figure*}

\begin{figure}
	\includegraphics[width=\columnwidth]{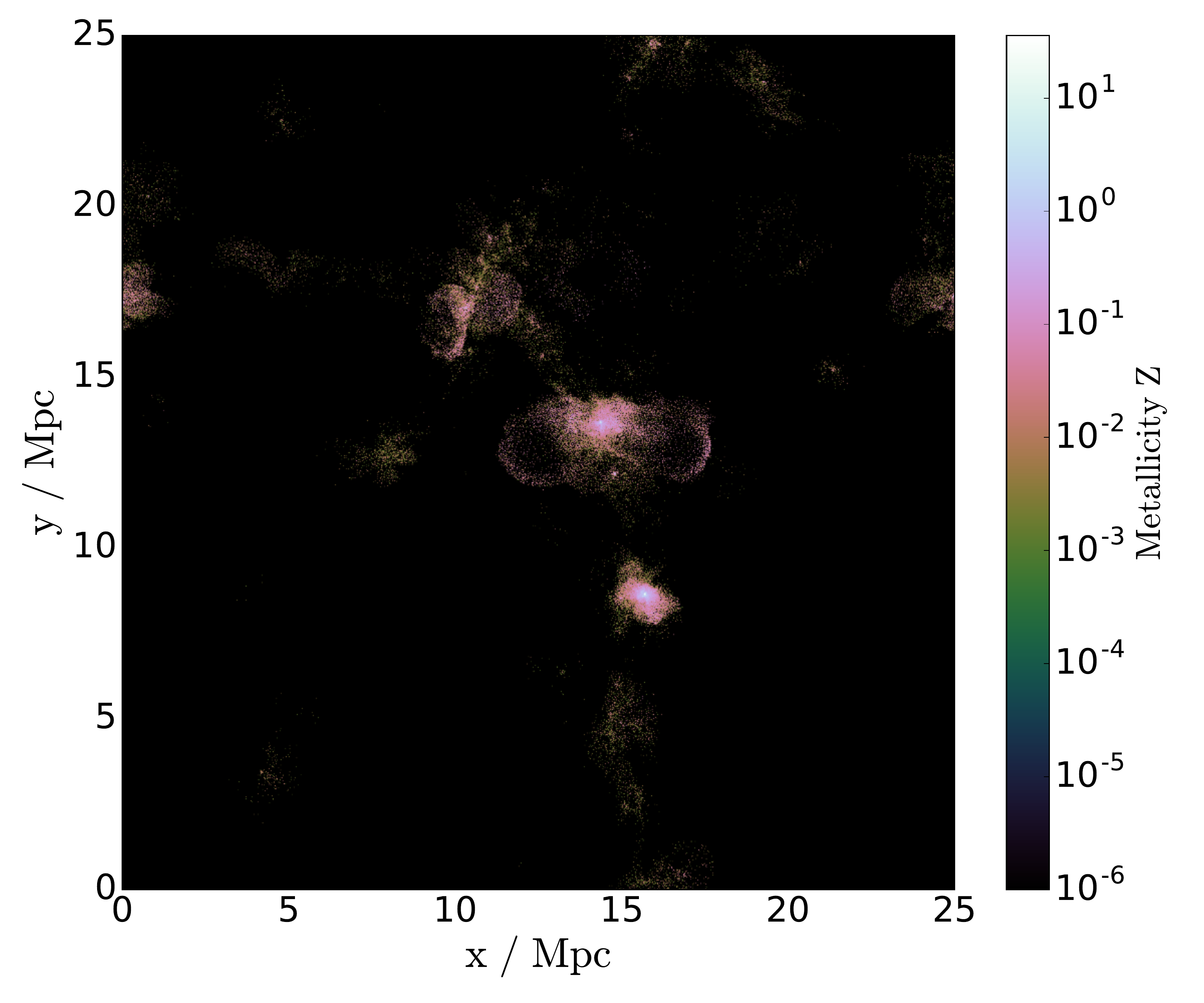}
    \caption{A 2.0 Mpc thick slice along the z axis of Ref-L0025N0376, showing the distribution of gas particles.
     A colourmap with a logarithmic contrast shows the metallicity $Z$ of these particles.
     Regions with $\rm{Z} = 10^{-6}$ have particles with zero metallicity. By comparing with
     Fig. \ref{Figure: Ref vs No Feedback}, it can be noticed that particles that are
     expelled from dense regions in the form of winds have a high metallicity, suggesting
     that they correspond to processed material being blown-away from haloes via SNe or AGN feedback.}
    \label{Figure: Metallicity}
    \centering
\end{figure}

As a motivation for the following sections that will focus on void statistics, we make use of the small EAGLE simulations to visually explore the effects of baryons on the large-scale distribution of matter. In Fig.~\ref{Figure: Gas vs Dark matter} we compare the distribution of gas and dark matter particles (left and right panels, respectively) in Ref-L0025N0376 at $z = 0.0$. The distribution of baryons looks more diffuse, and the high density regions are less concentrated than in the dark matter distribution. This could be due to different physical mechanisms implemented in the subgrid models of EAGLE, such as feedback from supernovae events, AGN feedback and gas cooling, all of which inject or remove energy from the gas and affect its spatial distribution. This feature has also been observed by \citet{2016MNRAS.457.3024H} in the Illustris cosmological simulation \citep{2014MNRAS.444.1518V}.

In Fig. \ref{Figure: Ref vs No Feedback} we compare the distribution of gas particles in Ref-L0025N0376 (left panel) and NoFeedback-L0025N0376 (right panel) at $z = 0.0$, where the colourmap now shows the temperature of gas particles. As mentioned previously, NoFeedback-L0025N0376 is a simulation that was run with a physical variation of the EAGLE code, in which stellar evolution feedback and AGN feedback have been supressed. Although the large-scale gas distribution is similar in both simulations, some clear signatures of baryon effects can be spotted in Ref-L0025N0376, which shows presence of hot gas in the surroundings of high density regions, as can be seen in the centre of the simulation. The fact that this supply of hot gas is not present in NoFeedback-L0025N0376 suggests that this gas comes from winds expelled from galaxies due to supernovae and AGN feedback.

Fig. \ref{Figure: Metallicity} shows the metallicity of gas particles in Ref-L0025N0376. It can be noticed that regions with high metallicity match with regions of high temperature in Fig. \ref{Figure: Ref vs No Feedback}, and it is interesting to see that the hot gas winds spotted in Fig. \ref{Figure: Ref vs No Feedback} show a large metal content, which confirms the picture that this gas corresponds to material that was processed in galaxies. These figures indicate that baryon effects trigger a complex and rich exchange of material between galaxies and the circumgalactic/intergalactic media.

In principle, the effects mentioned above could have an impact on void properties. The most intuitive way in which this could happen is when identifying voids using the mass field in the simulations. If feedback mechanisms inject mass in regions that would otherwise be empty, the properties of the voids identified near these regions could change. The voids could either shrink or have their centres displaced, and in a more extreme case a void could disappear from a particular region of the simulation if feedback mechanisms pollute that region with enough mass.

Baryon effects could also affect the properties of voids identified using subhalo tracers. As can be seen in Fig.  \ref{Figure: Ref vs No Feedback} and Fig. \ref{Figure: Metallicity}, feedback mechanisms can remove matter from galaxies and expell it to the surrounding medium. The most direct consequence of this would be a change in the mass of subhaloes that undergo these feedback events. Since we select the subhalo tracers using a stellar mass and a total mass cut, a change in the subhalo mass function could result in a population of voids with different properties.

An additional effect that could be present in the analysis of voids identified using subhalo tracers is the scatter in the stellar mass-subhalo mass relation, which naturally arises in a hydrodynamical simulation.  Notice that this is the only baryon effect that would also be present to some degree in a semi-analytic galaxy catalog.

\section{Void abundance} \label{Section: Void abundance}

\begin{table}
	\caption{A summary of some basic quantities associated to 
	void catalogs obtained in Ref-L0100N1504
	and DM-L0100N1504 (top and bottom tables, respectively) at $z=0.0$
	The first column shows the type of tracer that was used for void identification:
	subhaloes selected by their stellar mass, subhaloes selected by their total mass,
	or the mass field (individual particles).
	The remaining columns, from left to right, show the total number of voids,
	maximum, minimum and mean void radius for each sample.} 
	\label{Table: Void catalog}
	\begin{tabular}{rrrrr}
		\multicolumn{1}{|r|}{Ref-L0100N1504}\\
		\hline
		Void sample & $N_{\rm{void}}$ & $r_{\rm{max}}$ & $r_{\rm{min}}$  & $r_{\rm{mean}}$ \\
		& & $(\rm{Mpc})$ & $(\rm{Mpc})$ & $(\rm{Mpc})$\\
		\hline
		\textbf{40 per cent overlap}\\
		$\rm{M_{stel}}$ subhalo selection & 709 & 24.3 & 4.9 & 7.0 \\
		$\rm{M_{tot}}$ subhalo selection & 708 & 21.4 & 4.9 & 6.7\\
		Mass field & 7695 & 18.3 & 1.4 & 2.5\\
		\hline
		\textbf{30 per cent overlap}\\
		$\rm{M_{stel}}$ subhalo selection & 485 & 24.3 & 4.9 & 7.1\\
		$\rm{M_{tot}}$ subhalo selection & 470 & 21.4 & 4.9 & 6.8\\
		Mass field & 5680 & 18.3 & 1.4 & 2.5\\
		\hline
		\textbf{20 per cent overlap}\\
		$\rm{M_{stel}}$ subhalo selection & 341 & 24.3 & 4.9 & 7.1\\
		$\rm{M_{tot}}$ subhalo selection & 330 & 21.4 & 4.9 & 6.9\\
		Mass field & 4255 & 18.3 & 1.4 & 2.5\\
		\hline
		\\
		\multicolumn{1}{|r|}{DM-L0100N1504} \\
		\hline
		Void sample & $N_{\rm{void}}$ & $r_{\rm{min}}$ & $r_{\rm{max}}$  & $r_{\rm{mean}}$ \\
		& & $(\rm{Mpc})$ & $(\rm{Mpc})$ & $(\rm{Mpc})$\\
		\hline
		\textbf{40 per cent overlap}\\
		$\rm{M_{tot}}$ subhalo selection & 693 & 21.0 & 4.9 & 6.7 \\
		Mass field & 9536 & 18.6 & 1.4 & 2.4\\
		\hline
		\textbf{30 per cent overlap}\\
		$\rm{M_{tot}}$ subhalo selection & 466 & 21.0 & 4.9 & 6.8 \\
		Mass field & 6785 & 18.6 & 1.4 & 2.4\\
		\hline
		\textbf{20 per cent overlap}\\
		$\rm{M_{tot}}$ subhalo selection & 346 & 21.0 & 4.9 & 6.8\\
		Mass field & 5021 & 18.6 & 1.4 & 2.4\\
		\hline
	\end{tabular}
\end{table}

\begin{figure*}
	\includegraphics[width=\textwidth]{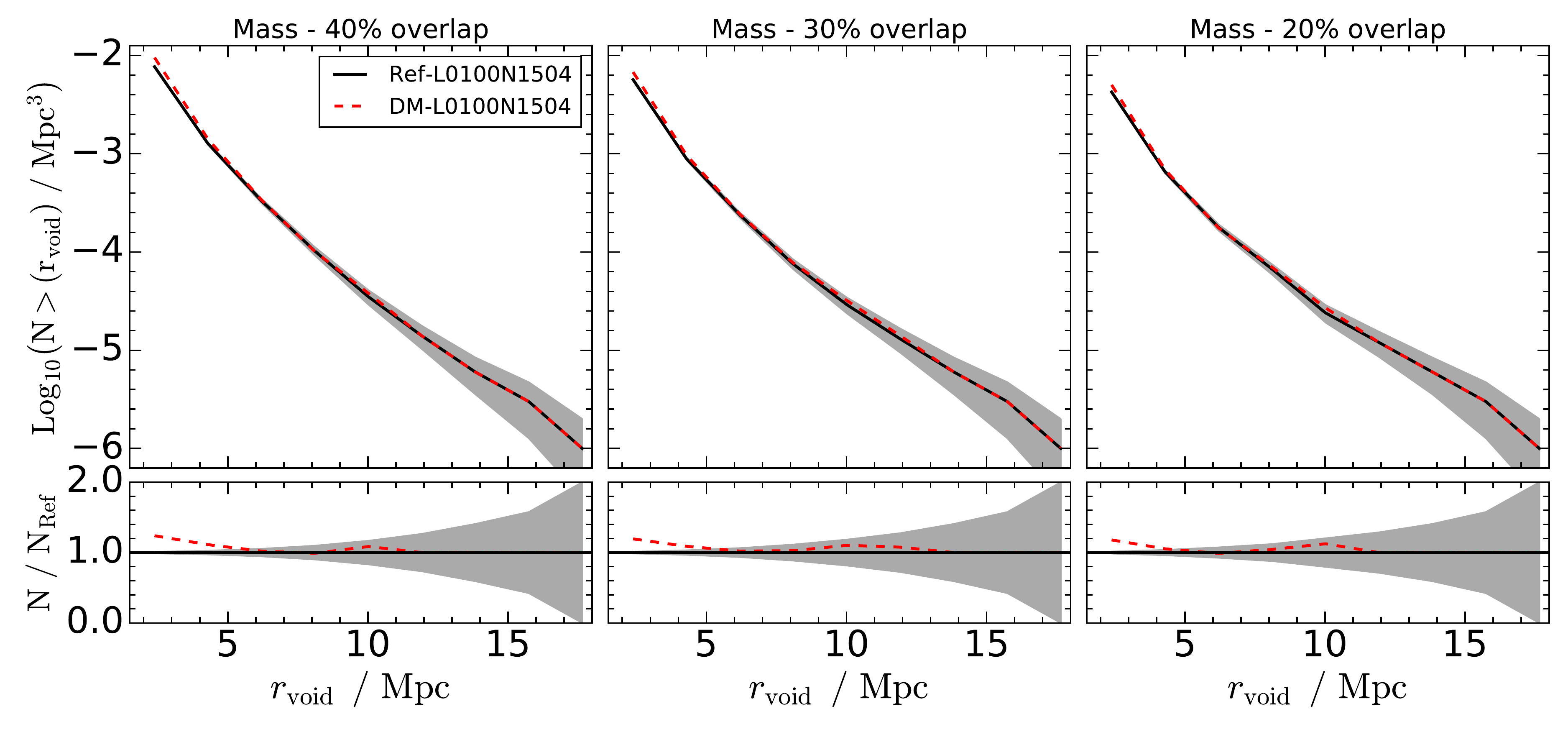}
    \caption{Upper panels: The cumulative distribution of sizes for voids identified with mass in Ref-L0100N1504 (black solid line),
    and DM-L0100N1504 (red dashed line) at z = 0.0
    Lower panels: The ratio between each of the lines in the upper panel and the black solid line.
    The left, middle and right panels show three independent void samples in which two adjacent voids can only have centres separated
    by more than 40, 30 and 20 per cent of the sum of their radii, respectively. 
    DM-L0100N1504 produces a larger amount of voids with $1.5 < r_{\rm{void}} < 5\ \rm{Mpc}$ than Ref-L0100N1504.
    The action of feedback mechanisms (SNe and/or AGN) in Ref-L010015N04 polutes under-dense regions with processed material,
    supressing the number of small voids identified in this simulation.
    }
    \label{Figure: Void abundance - mass field}
    \centering
\end{figure*}

\begin{figure*}
	\includegraphics[width=\textwidth]{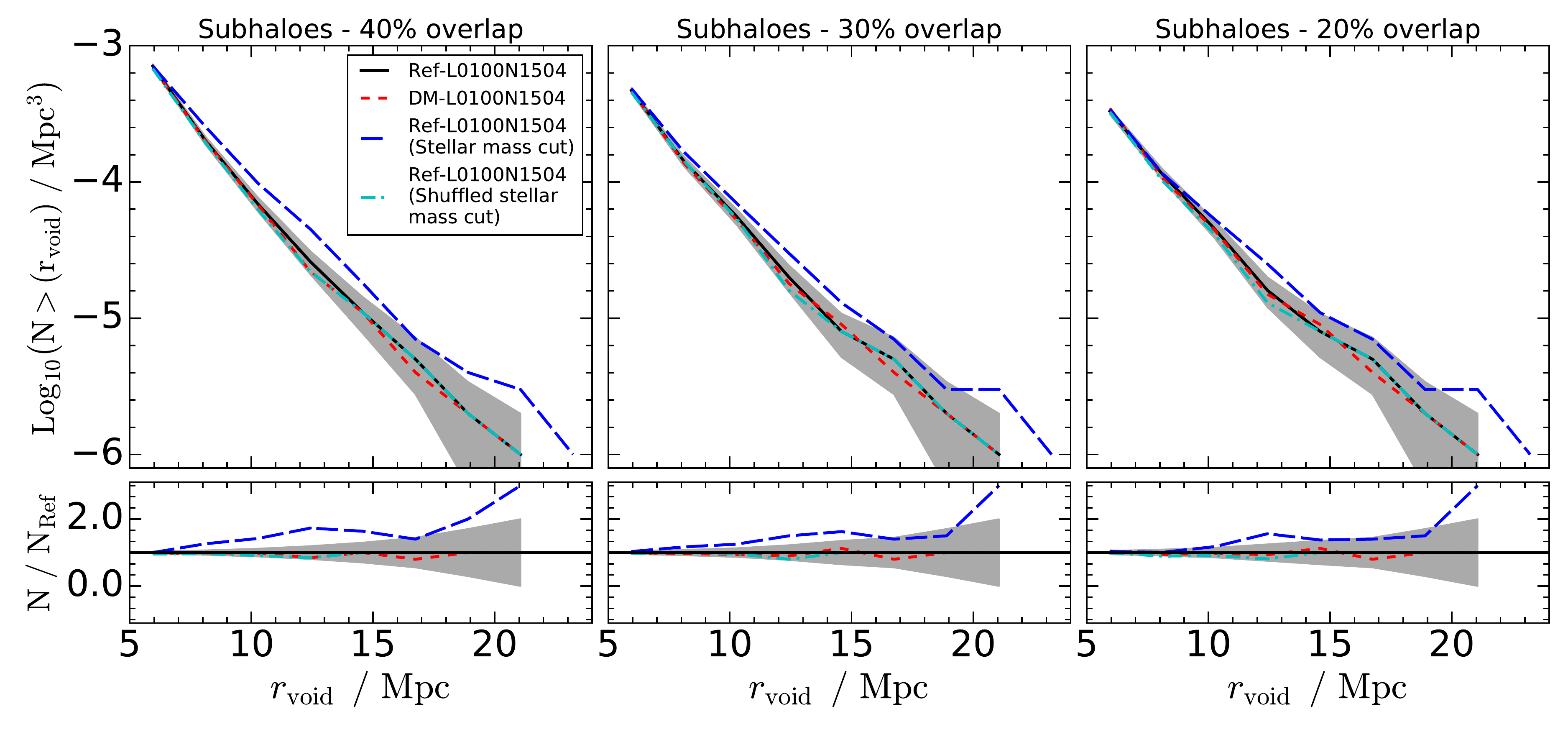}
    \caption{Upper panels: The cumulative distribution of sizes for voids identified with subhaloes selected by their total mass
    in Ref-L0100N1504 (black solid line), by their stellar mass in Ref-L0100N1504 (blue, long-dashed line), by their stellar mass
    in Ref-L100N1504, but after shuffling the stellar masses of all subhaloes in the simulation (cyan, dot-dashed line) and
    by their total mass in DM-L0100N1504 (red dashed line). All measurents correspond to z = 0.0
    Lower panels: The ratio between each of the lines in the upper panel and the black solid line.
    The left, middle and right panels show three independent void samples in which two adjacent
    voids can only have centres separated by more than 40, 30 and 20 per cent of the
    um of their radii, respectively. There are no significant differences between the reference
    and the DM-only models when the same total mass selection cut is used to find voids. However, voids
    traced by subhaloes selected by their stellar mass are systematically larger than
    those produced by a total mass cut.}
    \label{Figure: Void abundance - subhaloes}
    \centering
\end{figure*}

As mentioned in Section \ref{Subsubsection: Finding voids in EAGLE}, for each simulation we obtain voids identified in the mass  and the subhalo fields. For the latter case, two different selection cuts were used in Ref-L0100N1504): a stellar mass ($\rm{M_{stel}}$) and a total mass ($\rm{M_{tot}}$) cut.

\subsection{Voids identified using the mass field} \label{Subsection: Void abundance - mass field}

In Fig. \ref{Figure: Void abundance - mass field} we show the abundance of voids identified using the mass field in snapshots of Ref-L0100N1504 (black solid line) and DM-L0100N1504 (red dashed line) at $z=0.0$, adopting 10 bins in void radius.

The total number of voids is larger when more overlap is allowed between adjacent voids. This is expected, as when two voids share more volume than what is allowed by this parameter, only the largest one is kept in the catalog. The abundance of voids with 40 per cent overlap is higher than the catalogs with 30 and 20 per cent of overlap by a factor of 1.5 and 2.3, respectively.

When considering void catalogs with 40 per cent of overlap, a quick look at Table \ref{Table: Void catalog} reveals that 9536 voids were identified in the dark matter-only simulation and 7695 in the simulation with baryonic physics. This corresponds to roughly a 24 per cent difference in abundance, and similar values are found for the samples using 30 and 20 per cent of overlap. The dark matter-only simulation produces a larger number of voids with $1.5 < r_{\rm{void}} < 5\ \rm{Mpc}$, which could be a hint of baryonic processes adding (or retaining) mass in regions that would otherwise be more under-dense if these mechanisms were not present. DM-L0100N1504 also seems to produce a larger quantity of voids with $r_{\rm{void}} \sim 10\ \rm{Mpc}$, although the associated statistical uncertainties prevent us from drawing robust conclusions about this difference. A larger simulation volume is needed in order to study this effect with higher significance.

In principle, an injection of mass into an under-dense region would tend to shrink the void if we keep the centre fixed, since the integrated density criteria about that centre would be satisfied at a smaller radius. We must also remember that our algorithm imposes a minimum void radius cut. The motivation behind that threshold is that below some point, spherical regions cannot longer be resolved. If these void regions shrink below the minimum radius adopted, they are removed from the void catalog, which could explain the difference in the total number of voids between Ref-L0100N1504 and DM-L0100N1504.

Another possible scenario is that feedback mechanisms produce a displacement of the centre of the void. We remind the reader that our algorithm grows spheres about similar centres in the simulation, until the $\Delta = -0.8$ density threshold is satisfied. The overlapping criteria ensures that if two spheres have similar centres but different radii, only the largest sphere is identified as a void.  The location and size of this largest sphere can change depending on how these feedback mechanisms are affecting the environment, and a change in the position of a void can also affect smaller voids that surround it. Under this scenario, we could have a case in which a void in DM-L0100N1504 grows very large and the overlapping criteria deletes some of the smaller voids surrounding it. In Ref-L01001N504, this void could be smaller, which could result in the survival of the smallest voids that were deleted in the DM-only case, thus producing a difference in void abundance as seen in Fig. \ref{Figure: Void abundance - mass field}.

\subsection{Voids identified using subhaloes} \label{Subsection: Void abundance - subhaloes}

In the upper panels of Fig. \ref{Figure: Void abundance - subhaloes} we show the cumulative distribution of sizes for voids identified using subhaloes in snapshots of Ref-L0100N1504 and DM-L0100N1504 at $z=0.0$, using 10 bins in void radius. The black solid line shows voids traced by subhaloes selected by their total mass in Ref-L100N1504. The long-dashed blue line shows to voids traced by subhaloes selected by their stellar mass in Ref-L0100N1504. The red dashed line shows voids identified by subhaloes selected by their total mass in DM-L0100N1504. The left, central and right panel panel show void samples where two adjacent spheres cannot have centres closer than 40, 30 and 20 per cent of the sum their radii, respectively. The lower subpanels show the ratio between each curve and the black solid line. Grey regions show $1 \sigma$ Poisson errors for the black solid line.

Here the sample with 40 per cent of overlap has roughly twice as many voids as the 20 per cent sample, and nearly 1.5 times more than the 30 per cent sample, similar to what is found using the mass field.

By comparing the black and red lines, we see that differences in the abundance of voids in Ref-L0100N1504 and DM-L0100N1504 are consistent within the errors when subhaloes are selected by their total mass in the simulation with baryonic physics. However, the blue line shows that voids traced by subhaloes selected by their stellar mass are slightly larger than those traced by total mass-selected subhaloes, and even though the total number of voids is roughly equal in all cases, they show a over-abundance of more than 30 per cent for $r_{\rm{void}} > 10\ \rm{Mpc}$ for the sample with 40 per cent of overlap. This difference is smaller when less overlap is allowed between voids, which can also be noticed by looking at Table \ref{Table: Void catalog}, which summarizes the values for the total number of voids, the maximum, minimum and mean void radius for each sample. It is, however, interesting to see that the trends in Fig. \ref{Figure: Void abundance - subhaloes} are similar regardless of the value of the overlap parameter.

As shown by previous theoretical studies, there is a scatter in the relation between the host subhalo mass and the stellar mass of galaxies (e.g. \citealt{2015MNRAS.452.1861C, 2016MNRAS.461.3457G}). For instance, a subhalo with low stellar mass can still contain a substantial amount of dark matter. This implies that selecting subhaloes in EAGLE by their stellar or their total mass can produce a catalog of objects with different properties. If these two subhalo samples have a different spatial distribution, they might trace the cosmic web differently, which could translate in voids with different properties. To explore this in further detail, we construct a subhalo subsample in which we shuffle the stellar masses of all the subhaloes in Ref-L0100N1504 in 40 bins of total mass. We then construct a new subhalo sample selected by stellar mass with the shuffled catalog. This way, we obtain a void population almost exactly equal to the one traced by total mass-selected subhaloes, as shown by the cyan dot-dashed line in Fig. \ref{Figure: Void abundance - subhaloes}. This indicates that the relation between total and stellar subhalo mass is driving the differences between the black and the blue curves in Fig. \ref{Figure: Void abundance - subhaloes}.

Abundance matching results show that the galaxy stellar mass-halo mass relation becomes tighter at high stellar masses ($\gtrsim 10^9\,\rm M_{\odot}$; e.g. \citealt{2016MNRAS.456.1459M, 2013ApJ...770...57B, 2010ApJ...710..903M}). We attempted to adopt a higher stellar mass cut at the time of finding the voids. However, the statistics become restrictively poor, preventing us from reaching robust conclusions for the differences in the void populations caused by the structures used to trace the cosmic web. Such an analysis would require a larger simulated volume. Nevertheless, wide-area optical and near-IR redshift surveys, such as the SDSS and upcoming deep surveys, such as WAVES \citep{2016ASSP...42..205D}, are complete down to stellar masses even lower than we have adopted here, and thus we consider it pertinent to carefully investigate the results of our void finding algorithm using galaxies with stellar masses $>10^8\,\rm M_{\odot}$.

As discussed above, the fact that voids are larger when traced by subhaloes selected by their stellar mass suggests that these subhaloes have a different spatial distribution than those selected by their total mass. In Fig. \ref{Figure: Autocorrelations} of the Appendix we show the two-point autocorrelation function for the two sets of tracers, finding that subhaloes selected by their stellar mass are more strongly clustered than total mass-selected subhaloes. A stronger clustering signal indicates that these objects have a stronger bias with respect to the underlying mass distribution, as they live in higher density peaks of the cosmic web. This produces a catalog of voids with larger sizes. We remind the reader that our void finding algorithm grows spheres around prospective void centres until the integrated density profile is equal to 20 per cent of the mean number density of subhaloes in the simulation. If the subhalo tracers are preferentially located in higher density peaks, those spheres will be able to grow more before satisfying the density criteria, compared to the case where the spatial distribution of the tracers is more homogeneous.

Given that a significant contribution to the signal of clustering and weak lensing studies in future surveys will come from $z \sim 0.5$, we also obtain void catalogs for snapshots of Ref-L100N1504 at $z=0.5$ and $z=1.0$, in order to see if the differences in void abundance evolve as a function of time. Figures can be found in Appendix \ref{Appendix: Evolution with redshift}, where we show that the trends found at z = 0 are persistent at higher redshift.

\section{Void profiles} \label{Section: Void profiles}

\subsection{Void density profiles} \label{Subsection: Void density profiles}

\begin{figure*}
	\includegraphics[width=\textwidth]{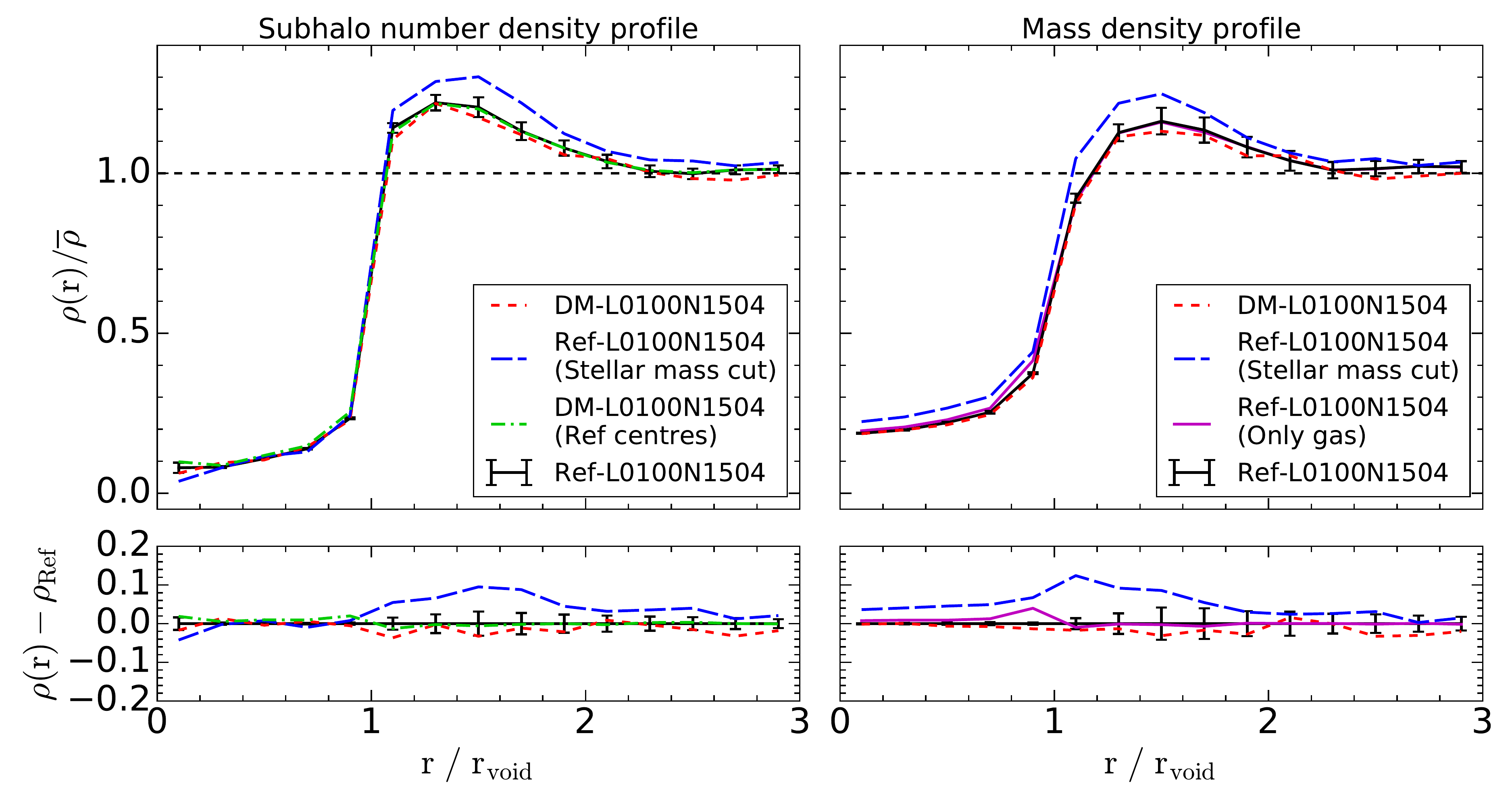}
    \caption{Upper panels: subhalo number (left) and mass (right) density profiles for voids traced by:
    subhaloes selected by their total mass in Ref-L0100N1504 (black solid line), subhaloes selected by
    their stellar mass in Ref-L0100N1504 (blue long-dashed line), subhaloes selected by their total mass
    in DM-L0100N1504 (red dashed line). The green dot-dashed line shows the density profile in
    DM-L0100N1504 for voids traced by total mass-selected subhaloes in Ref-L0100N1504. The magenta
    solid line in the right-hand side panel shows the mass density profile for total-mass selected subhaloes
    in Ref-L0100N1504, computed using only the gas particles in the simulation. All measurements 
    correspond to $z = 0$.  Lower panels: The absolute difference between each of the lines in the upper panels and the black
    solid curve of the upper panel. Error bars show the dispersion on the mean velocity profile for each radial bin. There are
    differences in the profiles of reference and DM-only model voids, even when the same total mass selection
    cut is applied for the tracers, this being more notorious at the peak of the profile. However, these
    differences are due to the different spatial distribution of the most massive subhaloes in the simulations.
    This effect is removed by using the same void centres in both simulations to calculate the profiles,
    as shown by the green dot-dashed line. Voids traced by stellar mass-selected subhaloes show a significantly
    higher peak, reflecting an overabundance of tracers near and within the walls of void regions.}
    \label{Figure: Density profiles}
    \centering
\end{figure*}

An important statistical tool to characterise void regions is their radial density profile. It is defined as the spherically averaged relative deviation of density around a void centre, compared to the mean value $\overline{\rho}$ across the simulation, $\rho_{r}\ /\ \overline{\rho}$. This profile can be computed in a differential or an integrated way.
 
To compute the differential void density profile, we measure the subhalo/mass density around a void centre in concentric radial bins of thickness $\delta r$. The density in such a shell can be computed as

\begin{equation}
\rho(r) = \frac{3}{4\pi} \sum_i \frac{w_i}{\left( r + \frac{\delta r}{2}\right)^3 - \left(r - \frac{\delta r}{2} \right)^3}\ \ ,
\end{equation}

\noindent where the sum is performed over all subhaloes/particles that fall within the radial bin that encompasses points between $r - \delta r$ and $r + \delta r$. The quantity $w_i$ corresponds to the weight assigned to a particular subhalo or particle in the computation. The void density profile is defined as the spherically averaged relative deviation of density around a void centre, compared to the mean value $\overline{\rho}$ across the simulation, $\rho_{r}\ /\ \overline{\rho}$.

When computing the subhalo density profile, $w_i = 1$, as every subhalo is assigned an equal weight in the calculation. When computing the mass density profile $w_i$ corresponds to the mass of the particle $i$, which has a  value of $9.70 \times 10^6\ \mathrm{M_{\odot}}$ if it is a dark matter particle, but can vary if it is a gas, star or black hole particle (gas particles initially weigh $1.81 \times 10^6\ \mathrm{M_{\odot}}$, but they can later gain mass via different processes; see \citet{2015MNRAS.446..521S} for more details). We re-scale each void-to-tracer distance by the void radius before stacking the profiles.

In a similar way, the integrated void density profile can be computed by measuring the density of subhaloes/mass in concentric spheres (instead of using shells) about a void centre. In this case, the integrated profile at a distance $r$ from the void centre would read

\begin{equation}
\Delta(r) = \frac{3}{4\pi} \sum_i \frac{w_i}{r^3}\ \ .
\end{equation}

In the upper panels of Fig. \ref{Figure: Density profiles} we show the differential subhalo (left) and mass (right) density profiles for voids identified using subhalo tracers in Ref-L0100N1504 and DM-L0100N1504. The black lines show voids identified using subhaloes selected by their total mass in Ref-L0100N1504. Voids identified using subhaloes selected by their stellar mass in Ref-L0100N1504 are shown by the blue long-dashed lines, while voids identified by subhaloes in DM-L100N1504 are represented by the red dashed lines. The lower panels show the absolute difference between each line and the black solid line. The error bars show the scatter around the mean for the black and blue lines. Error bars for the red line are not shown for clarity, but their sizes are similar to the other two, as we have explicitly checked. These profiles correspond to the void sample with 30 per cent of overlap. We have checked that results for the samples with 40 and 20 per cent of overlap show the same trends, with only minor differences in the size of the error bars due to the different number of voids.

The black and the red lines compare results in the simulations with and without baryonic physics, respectively, for voids identified using subhaloes tracers selected by their total mass. Looking at the subhalo number density profile (left-hand side panel), very small differences are observed between these two, although they are encompassed within the range of the statistical errors. It is important to stress, however, that it is expected that these two profiles are not identical, even though the initial conditions of both simulations are the same and the subhaloes were selected in the same way. \citet{2015MNRAS.451.1247S} used the EAGLE simulations to show that the mass of a subhalo in a hydrodynamical simulation can change substantially compared to its DM-only counterpart due to baryonic processes. Therefore, the most massive subhaloes in Ref-L0100N1504 and DM-L0100N1504 are located in different places in both simulations, and as a consequence, the voids traced by them can correspond to different structures. To remove this effect, we take the void centres found by total mass-selected subhaloes in Ref-L0100N1504, and compute their subhalo number density profile in DM-L0100N1504. This is represented by the green dot-dashed line, which falls right on top of the black solid line. As shown in the lower panel, these two curves have absolute differences very close to zero, which means that the main effect produced by baryons on the subhalo density profile of voids comes from a change in the subhalo mass function.

It is interesting to notice that voids traced by stellar-mass selected subhaloes show a significantly more pronounced over-dense ridge compared to the samples selected by total mass. This is in agreement with the discussion of the previous section, in which we commented on how selecting subhaloes by their stellar mass produces a distribution of tracers that are more strongly clustered. The fact that these subhaloes have a stronger clustering signal implies that they are preferentially located in over-dense environments, which shows up as an excess in amplitude at the void ridge when compared to voids traced by less-clustered subhaloes.

Although the subhalo number density profile is in principle something that can be observed in the real Universe through large-scale galaxy surveys, we must keep in mind that subhaloes form in over-dense peaks of the mass density field, and as such, they only provide a biased view of the mass distribution around and within voids. The right-hand side panel shows the mass density profile for the same voids discussed above. These density profiles were computed taking into account all the mass in the simulation, be it contained in gas, stars, dark matter or black hole particles. The overall shape of the profiles is similar to those computed using subhaloes, which means that subhaloes trace the mass distribution in voids to a reasonable extent. We have to keep in mind that these voids were selected using a subhalo number density threshold, so no explicit condition regarding the distribution of mass in these voids was imposed for their construction. In spite of this, some interesting differences with respect to the halo density profile can be noted. When looking at the mass distribution inside voids, we see that they are less empty of matter than what can be inferred by looking at the distribution of subhaloes alone. The lowest mass density is found at the void centres, reaching values as low as 20 per cent of the mean mass density in the simulation, while subhalo densities reach slightly lower values around 10 per cent of the mean density.

Another difference that is observed is that the discrepancy between voids identified by total-mass selected subhaloes and stellar mass-selected subhaloes now extends below $r\ /\ r_{\rm{void}} \sim 1$. This confirms the picture that these two void populations are distinct and they are being traced by different structures in the simulation.

Even though these mass profiles were computed using the mass contained both in baryons and dark matter, we must keep in mind that dark matter accounts for nearly 85 per cent of the total mass in the simulation. For this reason, the black and blue lines are mostly dominated by the dark matter distribution within and around voids in Ref-L0100N1504. Even though baryon effects such as galaxy feedback might also affect the distribution of dark matter, their most direct impact is expected to be seen in the distribution of gas. In order to study this we computed the mass density profile of voids using only gas particles. We do this for voids identified with total mass-selected subhaloes. This is shown by the magenta solid line in the right-hand side panel of Fig. \ref{Figure: Density profiles}. We find an excess of gas density below $r\ /\ r_{\rm~{void}} < 1$ compared to the distribution dominated by dark matter, which could be indicative of baryonic feedback processes polluting voids with pre-processed gas close to their walls. In this sense, the voids we find in the simulation contain a little more gas than dark matter, which is consistent with the picture of SNe and AGN winds injecting mass in void regions, as discussed in Section \ref{Section: Visual inspection of baryon effects}.

We have calculated the covariance matrices of these density profiles, and we do not find strong off-diagonal correlations among different bins, which means that we can compare the differences between lines and compare them with the error bars with confidence. More details can be found in Appendix \ref{Appendix: Covariance matrices}, where we also compute the reduced $X^2$ statistic to have a more quantitative measure of the statistical similarity between the profiles.

\subsection{Void velocity profiles} \label{Subsection: Void velocity profiles}

\begin{figure}
	\includegraphics[width=\columnwidth]{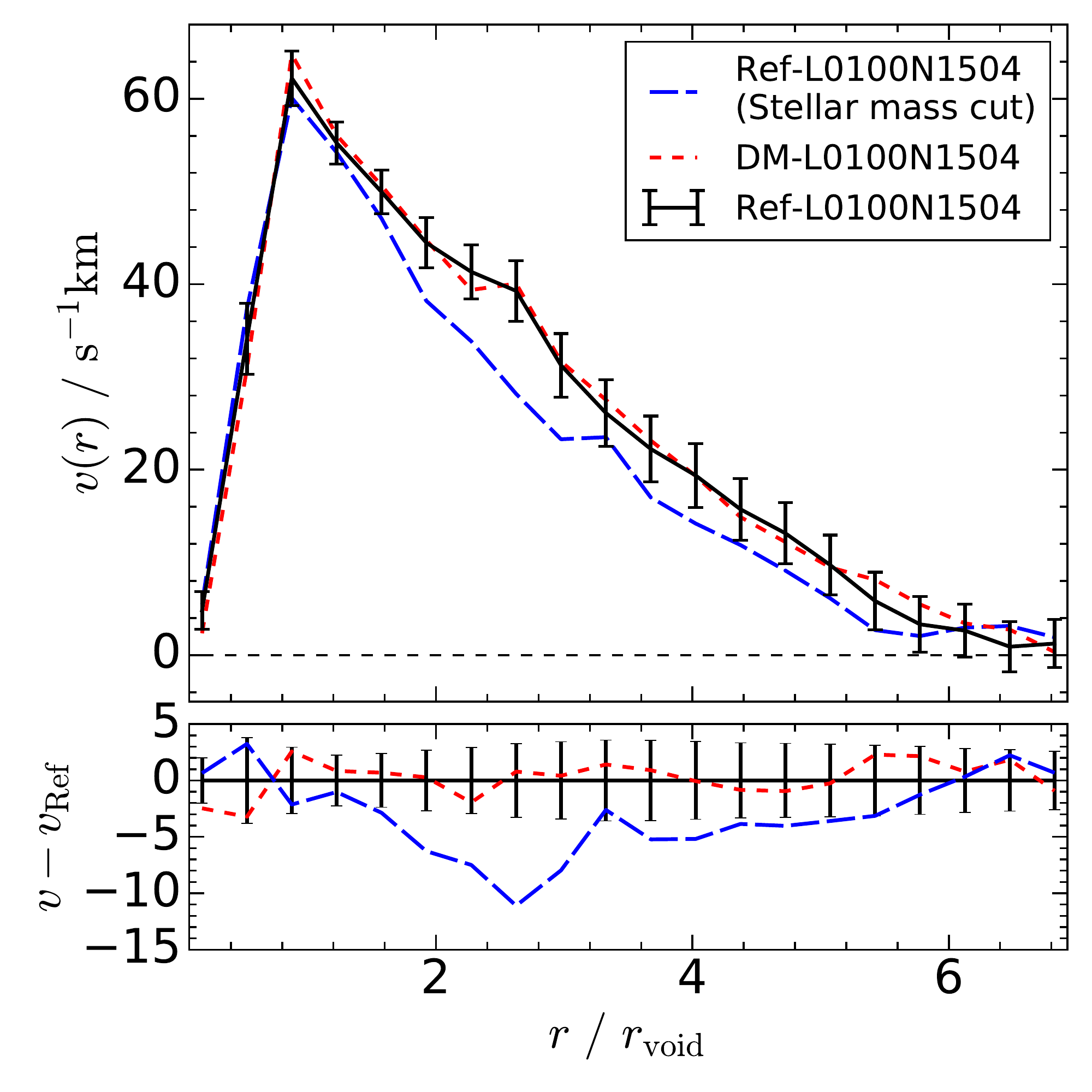}
    \caption{Upper panel: Radial velocities of subhaloes with respect to the centre of voids
     traced by: subhaloes selected by their total mass in Ref-L0100N1504 (black solid line),
     subhaloes selected by their stellar mass in Ref-L0100N1504 (blue long-dashed line),
     subhaloes selected by their total mass in DM-L0100N1504 (red dashed line). The green
     dot-dashed line shows the velocity profile in DM-L0100N1504 for voids traced by total
     mass-selected subhaloes in Ref-L0100N1504. All measurements correspond to $z=0$. 
     Lower panels: The absolute difference
     between each of the lines in the upper panels and the black solid curve. Error bars
     show the dispersion on the mean velocity profile for each radial bin. There are small
     differences between the reference and DM-only voids when the same total mass selection
     cut is applied for the tracers. This time, however, these differences cannot be reconciled
     by using the same centres in both simulations to compute the profiles (green dot-dashed line).}
    \label{Figure: Velocity profiles}
    \centering
\end{figure}

The large-scale structure of the Universe is far from static. Matter continuously flows in and out of void regions, and previous studies have suggested that there is a rich interchange of material between voids and other components of the cosmic web (e.g. \citet{2005MNRAS.363..977P, 2015MNRAS.451.1036C, 2016MNRAS.457.3024H}. The study of void velocity profiles is of great value for understanding void dynamics and evolution, as well as the relation between voids and the large-scale environment that surrounds them. It can also provide key information regarding redshift-space distortions around voids \citep{2016MNRAS.462.2465C}, which is essential for the interpretation of methods such as the AP test and the stacking of voids to get an ISW signal.

We can compute the radial velocities of subhaloes with respect to the centre of voids as:

\begin{align}
v(r) = \sum_i \frac{ \vec{v{_i}} \cdot \vec{d_i}}{\rm{N}(r) d_i}\ \ ,
\end{align}

\noindent where the sum is performed over all subhaloes that fall within the radial bin that encompasses points between $r - \delta r$ and $r + \delta r$. $\vec{v}_i$ corresponds to the velocity vector of subhalo $i$, while $\vec{d}_i$ denotes the position of the subhalo with respect to the centre of the void. $\rm{N}(r)$ corresponds to the number of subhaloes that fall within bin $r$. If $v > 0$, it means that there is a coherent outflow of subhaloes in that radial range (i.e. subhaloes evacuating from the void), while $v < 0$ corresponds to inflow velocities.

Fig. \ref{Figure: Velocity profiles} shows the velocity profiles for voids identified in Ref-L0100N1504 and DM-L0100N1504. We observe that there is a coherent outflow of subhaloes across a wide range of distances from the void centre. The outflow peaks near the void radius, consistent with the strong abundance of subhaloes in that region. The profile reaches velocities around 60 km/s, and at $r\ /\ r_{\rm{void}} > 1$ it decreases, converging to zero for larger distances.

Very small differences, within the size of the errorbars, are found between voids identified by subhaloes selected by their total mass in the simulations with and without baryonic physics (solid black and dashed red lines, respectively). This is in agreement with previous sections, which have shown that the abundance and density profiles of these voids are fairly similar.

The blue long-dashed line shows the velocity profile for voids traced by stellar mass-selected subhaloes.  The outflow velocities of these voids are consistent with the ones found by voids traced by total-mass selected subhaloes near $r\ /\ r_{\rm{void}} = 1$, but between $1 < r\ /\ r_{\rm{void}} < 5$ the velocities are lower by about $\rm{5 - 10\ km/s}$. We only show the error bars for the black line for clarity, but errors associated with the red and blue lines are of a similar magnitude.

It appears that the choice of tracer that is used for void identification has a strong impact not only on the size and density profile of a void, but also on its dynamics. In principle, this is something to be expected, since the dynamics of a void directly depend on the distribution of matter around it. Nevertheless, the trends observed in Fig.  \ref{Figure: Velocity profiles} need to be interpreted with care, given the small magnitudes of the velocity difference that is observed. Moreover, we have noticed that the differences between these profiles are sensitive to the overlap parameter of the void sample. The samples with 30 and 20 per cent of overlap have larger statistical errors and the differences between the black and blue line appear to be less significant. A larger simulated volume is needed to study this feature in greater detail.

One particular feature that calls our attention in these profiles is that we do not observe inflow velocities at any distance from the void centre. While some voids might be located in regions that are currently expanding, others might reside in over-dense environments that are collapsing or will eventually collapse, which would be translated in material flowing into the void. In the next section we discuss the existence of these two void populations in the context of EAGLE.

\section{Voids and their large-scale environment} \label{Section: Voids and their large-scale environment}

\begin{figure}
	\includegraphics[width=\columnwidth]{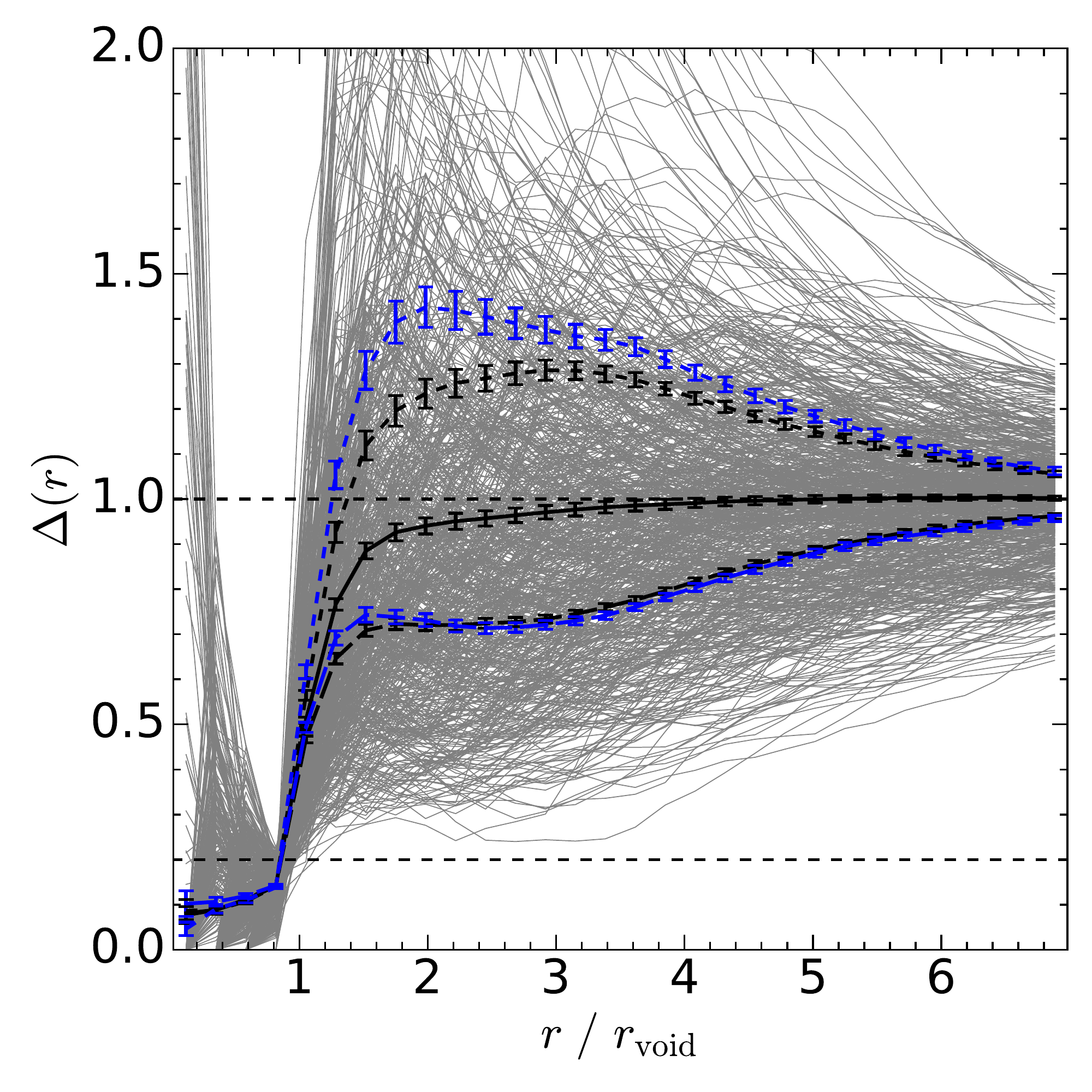}
    \caption{Integrated density profile for voids in Ref-L0100N1504 at $z=0.0$. The
    thin grey lines show individual profiles for voids identified using subhalo tracers
     selected by their total mass. The solid black line shows the average of these profiles.
    The dashed black lines above and below 1 show the average profile for voids classified
    as void-in-cloud and void-in-void, respectively, depending on whether their integrated density
    at $r\ /\ r_{\rm{void}} = 3$ is above or below 1. Blue dashed lines show the
    same results for voids identified using stellar-mass selected subhaloes.
    For reference, the horizontal dashed lines show the mean density of subhaloes and
    20 per cent of the mean density. Error bars show the scatter around the mean of each profile.}
    \label{Figure: Integrated density profiles}
    \centering
\end{figure}

\begin{figure}
	\includegraphics[width=\columnwidth]{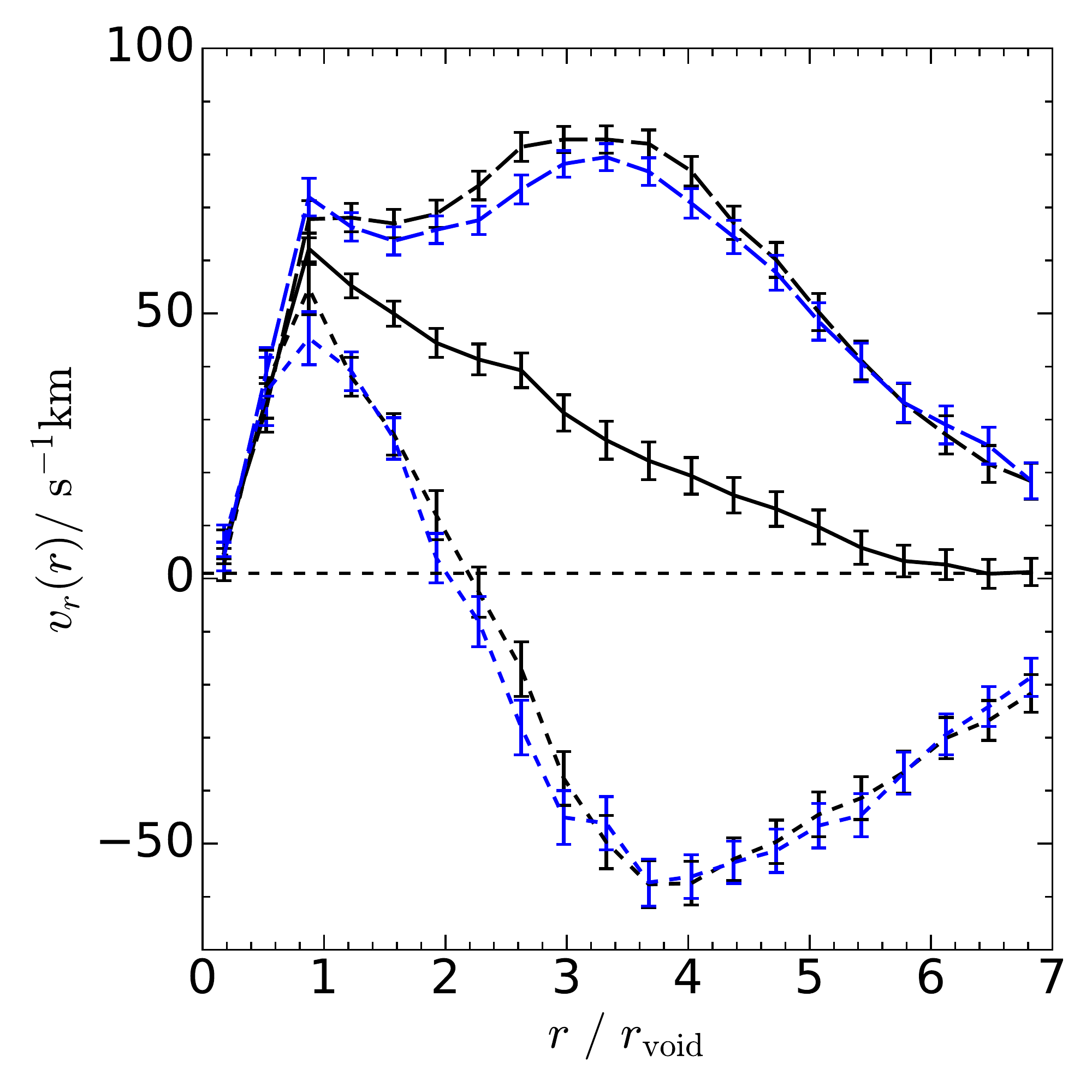}
    \caption{Radial subhalo velocity profile for voids in Ref-L0100N1504.
    The solid black line shows the mean profile for voids identified using subhalo tracers
    selected by their total mass. The long-dashed and short-dashed lines show the mean profiles
    for voids classified as void-in-cloud and void-in-void, respectively.
    Blue lines show results for voids identified using stellar mass-selected subhaloes.
    Error bars show the scatter around the mean of each profile.}
    \label{Figure: Velocity profiles - void-in-cloud and void-in-void}
    \centering
\end{figure}

As we have seen in previous sections, the mass or halo density in voids increases towards the void walls. The steepness of this increase holds important information about the dynamics of the void, as the total mass contained in their walls might over-compensate or under-compensate the mass that is missing inside the void. Under this context, some voids might be located in under-compensated regions that expand as the Universe evolves, while others might have over-compensating walls that will eventually make the void collapse. This has lead to a classification of voids into two distinct populations, void-in-cloud and void-in-void \citep{2004MNRAS.350..517S, 2013MNRAS.434.1435C, 2014ApJ...786..110C, 2014PhRvL.112y1302H}, depending on whether they are located in a large-scale environment that is collapsing or expanding, respectively.

\cite{2013MNRAS.434.1435C} classify voids into void-in-cloud and void-in-void depending on whether the accumulated overdensity at $r\ /\ r_{\rm{void}} = 3$ is above or below 0, respectively. \cite{2014PhRvL.112y1302H} and \cite{2015MNRAS.451.1036C} have shown that voids can also be separated in these different populations depending on their size. Voids with larger (effective) radii are more likely to be void-in-void, while void-in-cloud normally correspond to smaller voids.

We have chosen not to distinguish between these different types of voids in previous sections due to the restricted volume of the simulations we are analysing. We must keep in mind that the largest void in our sample is comparable in size to the smallest voids in the 1 Gpc / h simulation box used in \cite{2015MNRAS.451.1036C} and \cite{2014PhRvL.112y1302H}. The differentiation of these two void classes might depend on the large-scale modes present in any given simulation, and the size of EAGLE prohibits us from studying the large-radius end of the void size distribution. It is, nevertheless, still interesting to explore whether we can identify distinct void populations in EAGLE, as a way to improve our understanding between the void identification process and its relation to the size, resolution and tracer sampling of a simulation. Motivated by the results of previous sections, it is also important to determine the effect of baryons in the two void classes, in particular regarding the differences in void properties that arise as a result of using different subhalo samples to identify voids. We have to take this into account because due to simulation size, possible biases in our void samples could be present in the fraction of void-in-voids as a function of void radius.

In Fig. \ref{Figure: Integrated density profiles} we show integrated subhalo number density profiles for voids in Ref-L0100N1504. The thin grey lines show individual profiles for voids identified using subhalo tracers selected by their total mass. The mean of all these profiles is shown by the solid black line, with error bars that show the scatter around the mean. For reference, the upper and lower horizontal dashed lines show the mean density and 20 per cent of the mean density of subhaloes in the simulation. As can be noted, individual void profiles can differ significantly. Some voids appear to be embedded in highly over-dense environments, reaching values well above the mean density, while others have profiles that slowly increase towards the mean. It can be noticed that $\Delta=0.2$ at $r\ /\ r_{\rm{void}} = 1$, which corresponds to the density threshold imposed for void identification. High values of $\Delta$ are found near the centres of some voids. This is due to the presence of a small number of subhaloes in inner void regions that increase the integrated density well above the mean at such distances from the void centre. Some of these voids satisfy the density criteria at distances smaller than $r_{\rm{void}}$, but our algorithm guarantees that only the largest sphere satisfying the density criteria about any one centre is classified as a void.

When averaging over all curves, these voids show a profile that rises steeply out to $r\ /\ r_{\rm{void}} \sim 1$ and then converges slowly to the mean density for larger distances, as seen in the previous section. However, this profile arises after averaging over very distinct curves. Following \cite{2013MNRAS.434.1435C}, we separate voids in two different samples, depending on whether their integrated density is above or below 1 at $r\ /\ r_{\rm{void}} = 3$. The profile for these two samples are shown by the dashed and long-dashed black lines above and below the mean, respectively. These profiles resemble two distinct void populations: void-in-cloud are embedded in dense environments and have prominent ridges that over-compensate the lack of matter in their interiors. Void-in-void, on the other hand, show a profile that slowly increases towards the mean. They are under-compensated and do not feature an over-dense ridge, which suggests that these voids may be located in an expanding large-scale environment.

The dashed and long-dashed blue lines show these same profiles for voids identified using subhalo tracers selected by their stellar mass. While the void-in-void population seems to be rather unaffected by this different selection for the subhaloes, the void-in-cloud population shows a profile that reaches higher amplitudes between $1 < r\ /\ r_{\rm{void}} < 4$. This suggests that the differences observed in the abundance and density profiles of voids in the previous section come mainly from voids embedded in over-dense environments.

In Fig. \ref{Figure: Velocity profiles - void-in-cloud and void-in-void} we show the subhalo velocity profiles for these two void populations. We notice that voids embedded in over-dense environments show strong inflow velocities across a wide range of distances from the void centre. Voids that resemble the void-in-void population, on the other hand, exhibit strong outflow velocities, indicating that subhaloes are being evacuated from these regions. While these trends are consistent with what has been found in previous works (e.g. \citealt{Paz2013, 2013MNRAS.434.1435C}), the interpretation of these results must be taken with care due to the small volume of the simulation. It might well be the case that many of the voids that resemble void-in-void populations in this study may actually be located in regions that we would otherwise identify as over-dense if the volume of EAGLE was larger. It is, however, interesting to find that we can identify different kind of voids in such a small simulation box. The extent to which this difference is physical, and how this differentiation depends on the properties of the simulation will need to be answered in future works, as this area still remains largely unexplored.

\section{Conclusions and discussion} \label{Section: Conclusions and discussion}

We have analysed cosmic voids the EAGLE simulations, using subhalo tracers and the mass field for void identification. We have studied the impact of baryonic physics on void statistics by comparing void populations in the main hydrodynamical EAGLE simulation (Ref-L0100N1504) and its DM-only counterpart (DM-L0100N1504), which was run using the same initial conditions but only following the evolution of dark matter. EAGLE provides a unique opportunity to test the effects of baryons on void properties due to its high resolution and detailed treatment of baryonic physics, using subgrid models that follow star formation, radiative cooling, stellar feedback from massive stars, black hole growth and AGN feedback.

We define voids as spherical regions that have integrated density profiles equal to 20 per cent of the mean mass/subhalo density in the simulation. We have used a modified version of the spherical under-density void finder presented in \citet{2005MNRAS.363..977P}. When finding voids in the simulation with baryons using subhalo tracers, we used two different samples of subhaloes: The first one consists in a sample of subhaloes selected by stellar mass ($\rm{M_{stel} \ge 10^8\ M_{\odot}}$). The second sample consists in a set of subhaloes with the same number density, but selected by their total mass. In DM-L0100N1504, subhaloes are selected by their total mass, and the number density of subhaloes remains fixed and equal to the other two samples.

We have calculated the void abundance, density and velocity profiles for all the samples described above. We find effects originating from feedback, which expels gas into voids and, more importantly, from the stochasticity in the stellar mass that manages to form in subhaloes, that is from the scatter of the stellar mass-dark matter mass relation.

Our main results can be summarised as follows:

\begin{itemize}

\item DM-L0100N1504 produces about 24 per cent more voids than Ref-L0100N1504 when the mass field is used to identify voids. However, this difference is mainly comes from small voids with $1.5 < r_{\rm{void}} < 5\ \rm{Mpc}$ (Fig. \ref{Figure: Void abundance - mass field}). The action of feedback mechanisms from supernovae and AGN in Ref-L010015N04 pollutes void regions with hot gas, which causes some of these voids to shrink in size compared to their DM-only counterpart. If these regions shrink enough, they do not enter our sample of voids selected by our algorithm because their sizes fall below the minimum void size selection, which results in the hydrodynamical simulation having a lower overall abundance of voids than the DM-only simulation. This effect is also observed for voids with $r_{\rm{void}} \sim 10\ \rm{Mpc}$, although the associated statistical uncertainties are large. A bigger simulated volume is needed to test this effect with higher significance.

\item There are no significant differences in the abundance of voids identified using subhaloes in Ref-L0100N1504 and DM-L0100N1504 when subhaloes are selected by their total mass in both simulations. However, selecting subhaloes by their stellar mass tends to produce slightly larger voids (Fig. \ref{Figure: Void abundance - subhaloes}). This discrepancy arises because a stellar mass cut tends to select subhaloes that have a stronger bias with respect to the underlying matter distribution. These subhaloes are located in higher density peaks of the cosmic web than those selected by their total mass, which in the end results in larger voids. These differences are also seen for snapshots of EAGLE corresponding to $z=0.5$ and $z=1.0$

\item When subhaloes are selected by their total mass, small differences are seen in the subhalo density profiles of voids in Ref-L0100N1504 and DM-L0100N1504 (Fig. \ref{Figure: Density profiles}). These differences are mainly caused by the effect of galaxy feedback on the subhalo mass function of the simulations. When taking this effect into account by computing the density profile with subhaloes of DM-L0100N1504 for voids identified in Ref-L0100N1504, the profiles are almost identical. When subhaloes are selected by their stellar mass, voids show density profiles with a more pronounced peak, reflecting an overabundance of subhaloes around void walls. This confirms the differences found in void abundance, namely voids traced by galaxies selected by their stellar mass are larger due to the fact that these tracers are more strongly clustered than those selected by their total mass. These differences are also seen at $z=0.5$ and $z=1.0$.

\item The void mass density profiles show similar features as the profiles measured using subhaloes, meaning that subhaloes correctly capture the features of the mass distribution within and around voids with reasonable accuracy for a biased tracer of the density field. By measuring the void gas density profile we find an excess of gas near void walls, suggestive of the action of feedback mechanisms polluting voids with hot gas coming from galaxy winds.

\item No significant differences are found in the velocity profiles of voids in the simulation with baryons and its DM-only counterpart, if the subhaloes are selected by their total mass. For the case where subhaloes are selected by their stellar mass, subhaloes evacuate void regions at slightly lower (5 - 10 km/s) velocities than in the total mass case. The void velocity profiles appear to be very sensitive to the overlap that is allowed between voids in our algorithm, while the associated statistical uncertainties are still too large to establish a robust conclusion about this trend.

\item We are able to identify two different voids populations in EAGLE: voids embedded in under-dense large-scale environments that appear to be expanding, and voids in contracting dense environments. These resemble the void-in-void and void-in-cloud populations found in previous works. The effects of baryons appear to be more significant for the void-in-cloud, as these voids show distinct density and velocity profiles when the subhaloes by which they are traced are selected by their stellar or total mass. The fraction of void-in-void that we find is probably affected by the restricted number of large-scale modes that are present in these simulations. Nevertheless, this raises many questions about the relation between this void classification and the properties of the simulations in which these structures are identified.

\end{itemize}

It appears that void properties are well captured by DM-only simulations, with baryons only adding second-order effects, which are less important than those so far reported for alternative cosmologies.

When identifying voids in the mass field, we find that a DM-only simulation tends to produce larger voids than a hydrodynamical one due to the effects of galaxy feedback. While this is an interesting result that sheds light on the impact of baryonic processes on the large-scale distribution of matter, it does not directly have an impact on observational studies of voids identified in galaxy surveys. We do find, however, that care must be taken with the galaxy tracers that are used to find and characterise voids, since they can have a strong impact on the properties that are inferred. The different results from stellar mass and subhalo mass cuts indicate that this check needs to be done for each void search, mimicking the number density of the tracer sample. Moreover, the bias of the galaxy sample needs to be reproduced by the haloes used to identify voids in a mock sample. The fact that these effects are also present at higher redshifts in the simulations suggests that they should be considered when studying voids in current and future large-scale galaxy surveys, especially in the context of studies that aim to constrain cosmological models from void statistics. The results of this work are relevant in particular to spherical-based void finders, which are more suitable in the case of weak-lensing studies. It is worth noting, however, that it is difficult to predict whether the differences that we find in this work can be extrapolated to other void finders, especially to those that are not spherical-based. Nevertheless, this opens the possibility for future works to expand on this matter.

Given the size of the EAGLE simulation, we are not able to make predictions for voids larger than $\sim 25\ \rm{Mpc}$, which is comparable to voids identified in the main galaxy sample of SDSS. This is the best that can be done including all baryonic physics to date, but since we have found that the actual gas distribution is not as important as the galaxy population that is used to identify voids, in Paillas et al. (in prep) we will look at voids in different semi-analytic models of galaxy formation, which will be relevant to many more surveys. It is also worth noting that it is difficult to predict whether the differences that we find in this work can be extrapolated to other void finders. 

Considering the results obtained in this study, a suitable avenue for future works would be to combine DM-only simulations with abundance matching or semi-analytical models to populate the DM haloes with galaxies in order to characterise voids closely to what observers would do.

\section*{Acknowledgements}

We are grateful for the helpful comments and suggestions made by the anonymous referee. We thank Carlton Baugh and Joop Schaye for insightful discussions and comments. EP is supported by Proyecto Financiamiento BASAL PFB06. CL is funded by a Discovery Early Career Researcher Award (DE150100618). NP is supported by "Centro de Astronom\'ia y Tecnolog\'ias Afines" BASAL PFB-06 and by Fondecyt Regular 1150300. EP and NP acknowledge support for exchanges between PUC and Durham U. from the Newton-CONICYT fund project DPI20140114. Parts of this research were conducted by the Australian Research Council Centre of Excellence for All-sky Astrophysics (CAASTRO), through project number CE110001020. This work used the DiRAC Data Centric system at Durham University, operated by the Institute for Computational Cosmology on behalf of the STFC DiRAC HPC Facility (www.dirac.ac.uk). This equipment was funded by BIS National E-infrastructure capital grant ST/K00042X/1, STFC capital grants ST/H008519/1 and ST/K00087X/1, STFC DiRAC Operations grant ST/K003267/1 and Durham University. DiRAC is part of the National E-Infrastructure. Part of the analysis was done on the Geryon cluster at the Centre for Astro- Engineering UC. The Anillo ACT-86, FONDEQUIP AIC-57, and QUIMAL 130008 provided funding for several improvements to the Geryon cluster. This work was supported in part by the Science and Technology Facilities Council (grant number ST/F001166/1), European Research Council (grant numbers GA 267291 "Cosmiway") and CONICYT PIA Anillo ACT- 1417.




\bibliographystyle{mnras}
\bibliography{mnras_template.bib} 



\appendix \label{Appendix}

\section{Convergence tests} \label{Appendix: Convergence tests}

\begin{figure}
	\includegraphics[width=\columnwidth]{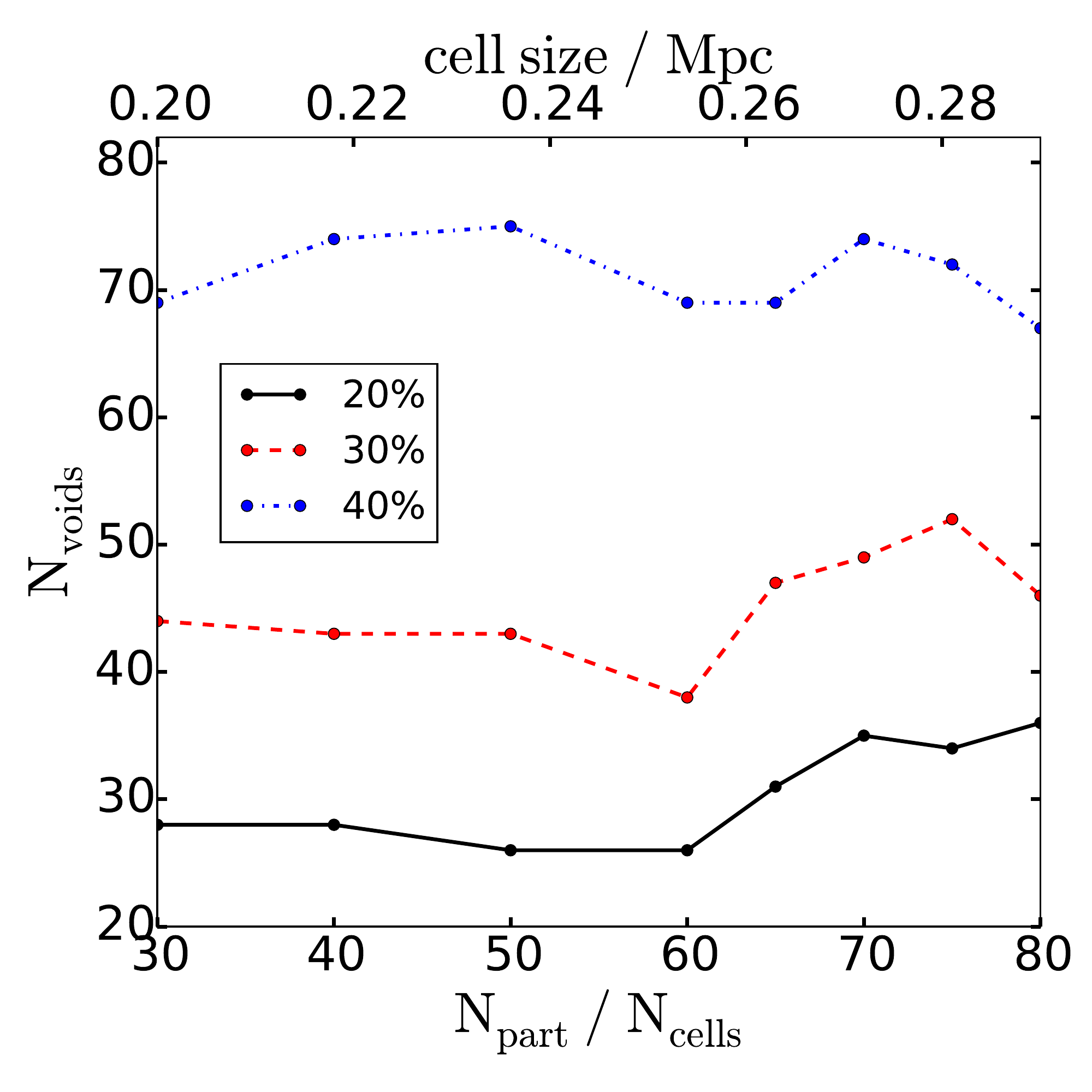}
    \caption{Number of voids traced by mass in Ref-L0100N1504 at $z=0$, as a function of the size
    of the cell used to smooth the density field in the first step of the void finding algorith 
    (Sec. \ref{Subsubsection: Void finding algorithm}). The lower horizontal axis shows the average number of particles
    that fall in each cell of the rectangular grid, while the upper horizontal axis shows the size
    of the cells in Mpc. The solid black, dashed red and dot-dashed blue lines show results
    for void samples with overlaps of 20, 30 and 40 per cent, respectively.
    These voids achieve convergence in their abundance, and we therefore use a cell size equal to 0.3 Mpc
    to identify voids traced by mass in Ref-L0100N1504.}
    \label{Figure: Convergence test}
    \centering
\end{figure}

As mentioned in Sec. \ref{Subsubsection: Void finding algorithm}, our void finding algorithm starts by constructing a rectangular grid and computing the number of particles (or haloes) that fall on each cell of the grid. Only empty cells are considered as prospective void centres. Naturally, decreasing the size of the grid cells (i.e. increasing the grid resolution) increases the number of prospective centres, but spurious void centres also appear more frequently. These spurious centres are eventually removed in further steps of the algorithm, but they increase the computational time of the void identification process. When identiying voids using subhalo tracers, this increase in computational time is negligible, since the number of subhaloes in EAGLE is small and the void finder can be run in a matter of seconds. In this case, we can increase the resolution of the grid a lot and we are still able to run the algorithm in a reasonably short time. When identifying voids in the mass field, however, the computational time for void identification is of the order of 40000 hours of CPU time due to the large number of particles in EAGLE, so we cannot arbitrarily increase the resolution of the grid. We are interested to know to what extent the identified void sample changes depending on the resolution of the grid. The expectation is that above a given resolution, the number of voids increases only due to spurious void centres, that are removed after applying steps (ii) and (iii) of our algorithm.

We run a convergence test in Ref-L0025N0376 to address this issue. Since this simulation has the same resolution and subgrid physics as Ref-L0100N1504, these tests are also applicable to void samples in the big simulation. Fig. \ref{Figure: Convergence test} shows the abundance of voids identified using the mass field in Ref-L0025N0376 as a function of the cell size of the grid used in the first step of the void finder. The solid black, dashed red and dot-dashed blue lines show results for void samples with overlaps of 20, 30 and 40 per cent, respectively. The lower horizontal axis shows the average number of particles that fall in each cell of the rectangular grid, while the upper horizontal axis shows the size of the cells in Mpc. The abundance of voids is not severely affected when reducing the size of the grid cells below $\sim 0.3\ \rm{Mpc}$. Moreover, we have explicitly examined the void size distribution for each of these samples, finding that the dispersion in the mean void radius is less than 2 per cent. Taking this into consideration, we choose to construct a grid with a cell resolution of $0.3\ \rm{Mpc}$ to search for voids in Ref-L0100N1504 using the mass field.

\section{Clustering measurements} \label{Appendix: Clustering measurements}

\begin{figure}
	\includegraphics[width=\columnwidth]{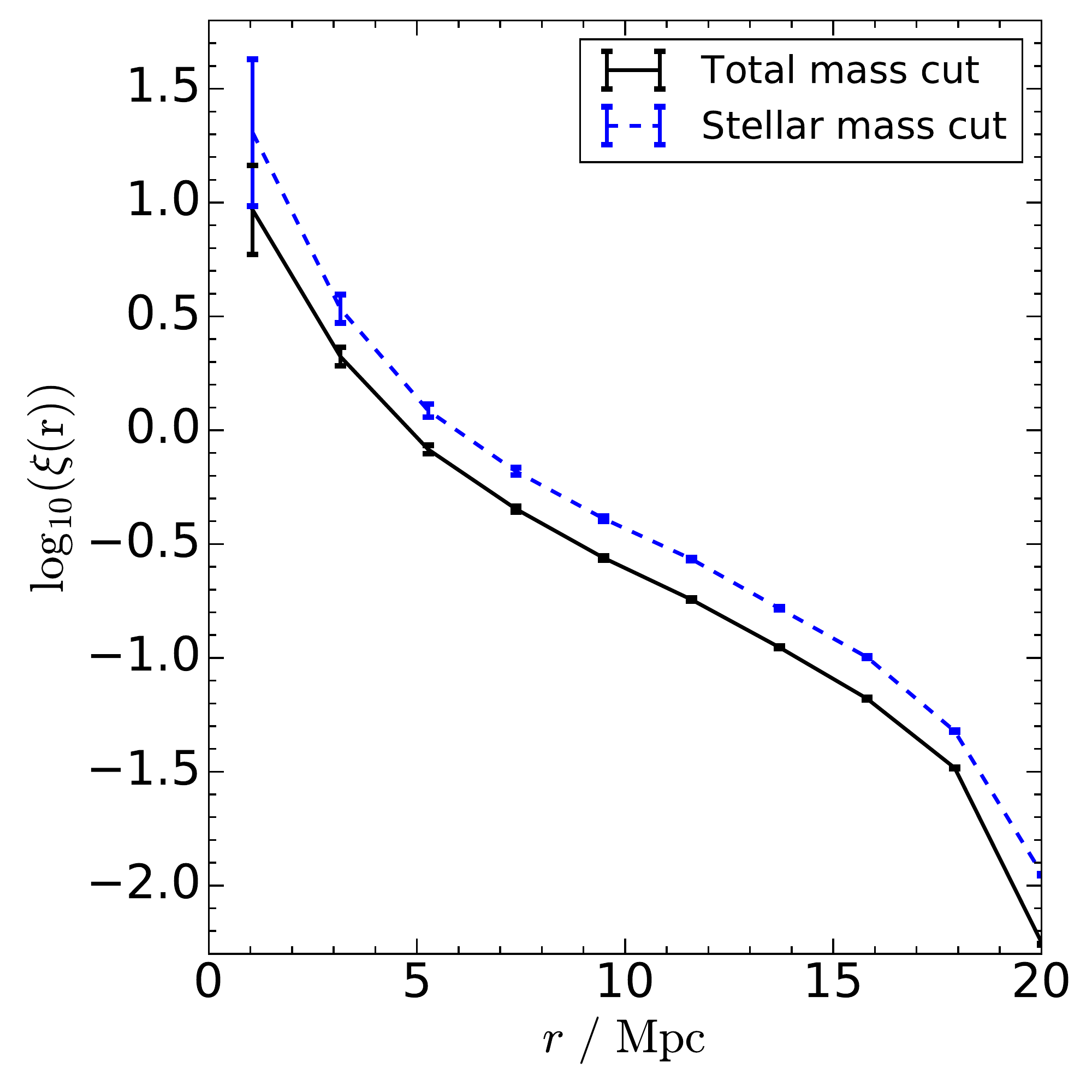}
    \caption{Two-point autocorrelation function of subhaloes used to trace voids Ref-L0100N1504 at $z = 0$.
    The black-solid and blue-dashed lines show measurements for subhaloes selected by their total or stellar 
    mass, respectively. Stellar mass-selected subhaloes show stronger clustering at all scales considered,
    which explains the increased size of the voids they trace.}
    \label{Figure: Autocorrelations}
    \centering
\end{figure}

The two-point correlation function (or simply the autocorrelation function) $\xi(r)$ is among the most widely used statistical tools to study the spatial distribution of galaxies
in the Universe. It measures the excess probability of finding a pair of points separated by a distance $r$,
compared to a random distribution \citep{1980lssu.book.....P}.

This function can be computed via the estimator \citep{1993ApJ...412...64L}

\begin{align}
\hat{\xi}_{\rm{LS}}(r) = \frac{DD - 2DR + RR}{RR}\ \ ,
\end{align}

\noindent where $DD$, $DR$, and $RR$ correspond to the number of data-data, data-random and random-random pairs in each separation bin, with a suitable normalization.

In Fig. \ref{Figure: Autocorrelations} we show an estimation of the two-point correlation function for the subhalo samples used to identify voids in Ref-L0100N1504 at $z=0.0$, calculated using 30 bins of distance between 0 and 20 Mpc. The black solid and blue dashed lines show measurements of the autocorrelation function for subhaloes that were selected by their total or stellar mass, respectively. Error bars show show bootstrap errors, computed using 40000 bootstrap re-samples. Each bootstrap re-sample is  constructed by sampling with replacement 40076 galaxies from the original set. For each of these re-samples, the autocorrelation function is computed, and the error bars are obtained from the standard deviation of the distribution of correlation functions. We observe that subhaloes selected by their stellar mass are more strongly clustered than those selected by their total mass over a wide range of scales. This suggests that the former are more likely to be located in denser environments of the cosmic web such as walls and filaments. In this scenario, voids traced by these subhaloes will be larger than voids traced by a more homogenoeus distribution of points, since the density criteria used for the definition of a void will be satisfied at a larger radius, as explained in Sec. \ref{Subsection: Void abundance - subhaloes}. This explains the increased size of voids traced by stellar mass-selected subhaloes seen in Fig. \ref{Figure: Void abundance - subhaloes}, as well as a higher peak of the density profile for these voids in Fig. \ref{Figure: Density profiles}.

\section{Evolution with redshift} \label{Appendix: Evolution with redshift}

\begin{figure}
 \includegraphics[width=\columnwidth]{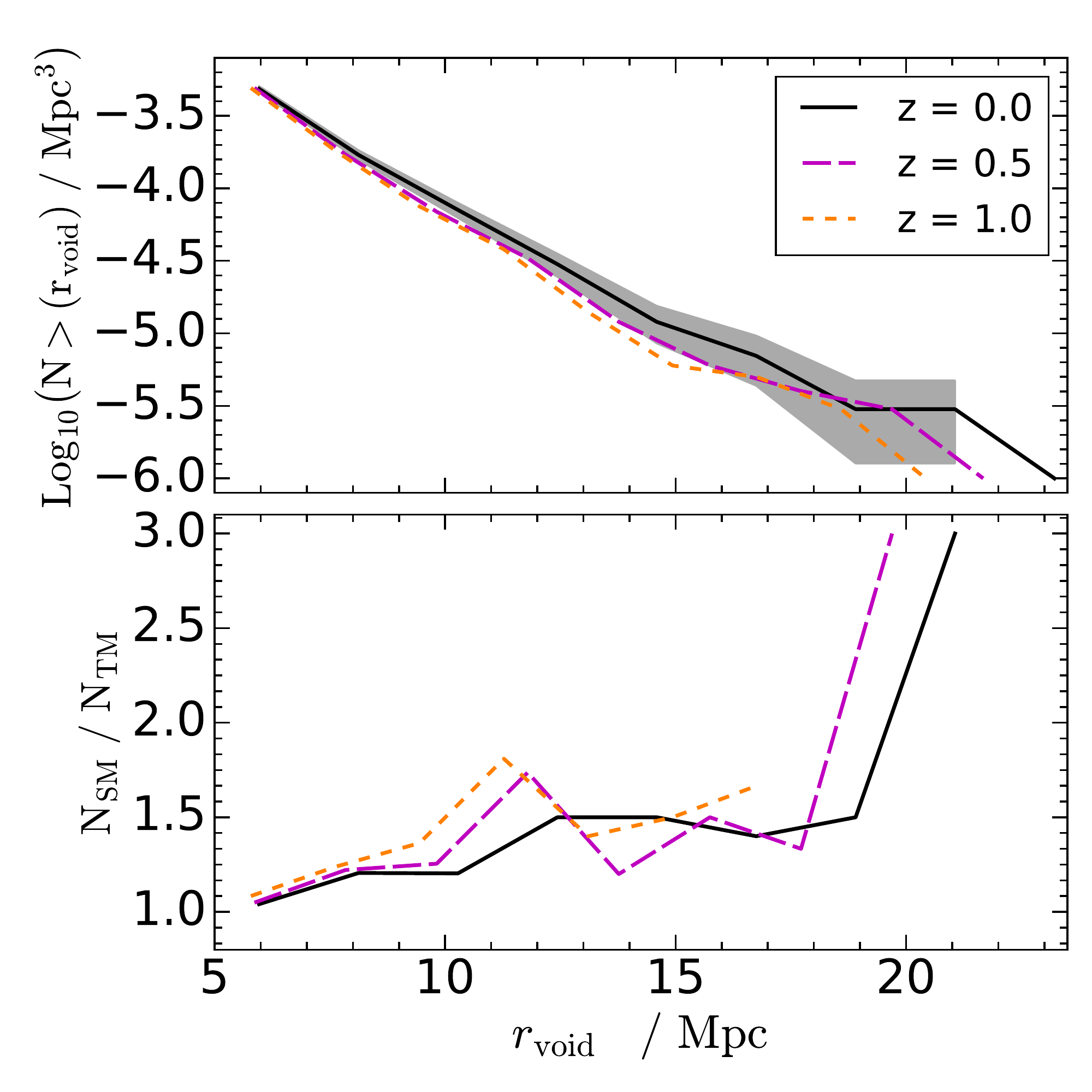}
    \caption{Top panel: cumulative distribution of sizes for voids traced by subhaloes selected by their stellar mass in
    Ref-L100N1504. Measurements for $z=0$, $z =0.5$ and $z=1.0$ are shown by the solid black, long-dashed magenta
    and dashed orange lines, respectively. While the total number of voids remains similar in all three cases, their size
    increases as a function of time as more matter gets accreted into filaments and haloes.
    Lower panel: abundance ratio between voids traced by subhaloes selected by their stellar/total mass in Ref-L0100N1504.
    Irrespectively of the redshift considered, voids traced by subhaloes selected by their stellar mass are larger
    than the other void samples.}
    \label{Figure: Void abundance - redshift}
    \centering
\end{figure}

\begin{figure}
 \includegraphics[width=\columnwidth]{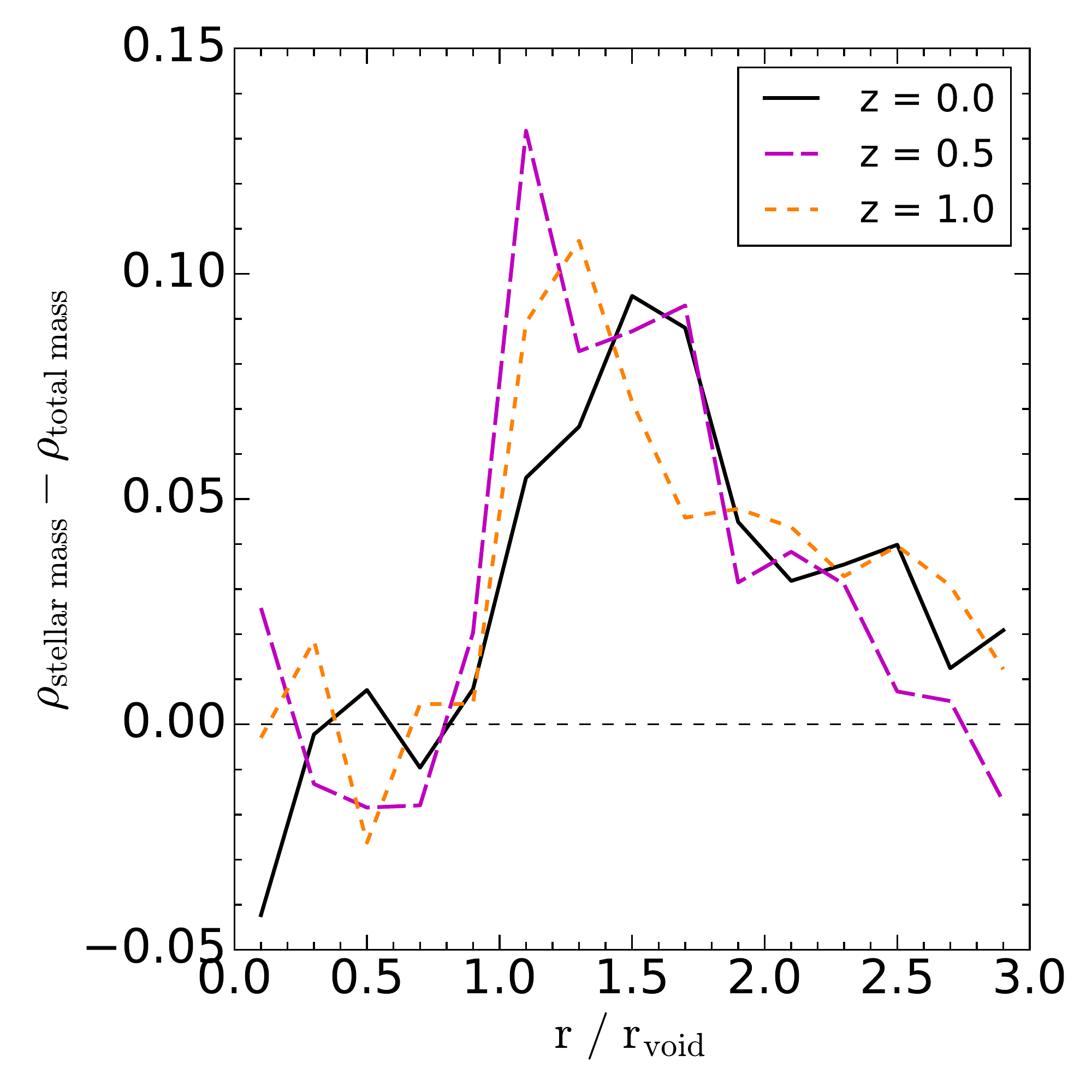}
    \caption{Absolute difference in the subhalo density profiles between voids traced by stellar mass-selected subhaloes
    and total mass-selected subhaloes. Measurements for $z=0$, $z =0.5$ and $z=1.0$ are shown by the solid black, long-dashed magenta
    and dashed orange lines, respectively. 
    Irrespectively of the redshift considered, voids traced by subhaloes selected by their
    stellar mass show a greater abundance of tracers around the void ridges. }
    \label{Figure: Density profiles - redshift}
    \centering
\end{figure}

In previous sections we have shown that when galaxies are used as tracers for voids, differences in the void abundance and void density profile arise, in part, due to how galaxies of a given mass populate haloes of a given mass and environment. From an observational point of view, it is important to consider that much of the signal from future large-scale surveys, such as LSST \citep{2009arXiv0912.0201L}, EUCLID \citep{2011arXiv1110.3193L} or eBOSS \citep{2016AJ....151...44D}, will come from galaxies at different redshift ranges. For example, over the time between $z \sim 0.6$ and $z = 0.0$, the relation between halo mass and stellar mass might have changed (e.g. \citealt{2013MNRAS.428.3121M}). Moreover, given that future large-scale clustering and weak-lensing studies will heavily rely on galaxies detected near $z \sim 0.5$, it is important to check whether our results change at earlier epochs in the simulation.

In Fig. \ref{Figure: Void abundance - redshift} we show the distribution of sizes for voids identified using subhalo tracers in snapshots of Ref-L0100N1504 at $z=0$, $z=0.5$ and $z=1.0$, shown by the solid black, long-dashed magenta and dashed orange lines, respectively. For better clarity, the upper panel only shows the abundance for voids traced by subhaloes selected by their stellar mass, while the lower panels show the ratio between the abundance of these voids and those traced by subhaloes selected by their total mass.

We observe that while the total number of voids is fairly similar in all the samples, the voids identified at higher redshift are somewhat smaller, although the associated uncertainties are still large. The interpretation of this difference in void size is not straightforward, as there are many mechanisms that interplay. These voids at higher redshift were identified using the same integrated density threshold $\Delta = -0.8$ as in the previous sections. Whether this threshold should be modified according to the growth of fluctuations is still a subject of debate in the field. Moreover, in Sec. \ref{Section: Voids and their large-scale environment} we showed that different void populations can be identified in EAGLE, some of them being embedded in large-scale environments that are expanding, while others are contracting. These and other factors might affect the distribution of void sizes at different redshifts. A detailed understanding of this evolution is outside the scope of this work, and a larger simulated volume is required to avoid possible biases in the determination of the large-scale environment of the largest voids in the sample.

Focusing on the lower panel of Fig. \ref{Figure: Void abundance - redshift}, we observe that selecting subhaloes by their stellar mass produces larger voids than the total mass selection, irrespectively of the redshift. This indicates that the differences observed in the void size distributions in Sec. \ref{Section: Void abundance} are persistent at $z = 0.5$ and $z=1.0$.

Fig. \ref{Figure: Density profiles - redshift} shows the absolute difference in the density profiles between voids traced by stellar mass-selected subhaloes and total-mass selected subhaloes in Ref-L0100N1504. Again we find at at the three redshifts considered, voids traced by subhaloes selected by their stellar mass always show density profiles with a higher peak, reflecting a greater abundance of subhaloes around and within the void ridges. A small shift in the radius at which we get the largest differences between the two samples above is obtained with redshift, but the measurements are too noisy to make a conclusive statement about this feature.

\section{Error estimations} \label{Appendix: Covariance matrices}

\begin{figure}
	\includegraphics[width=\columnwidth]{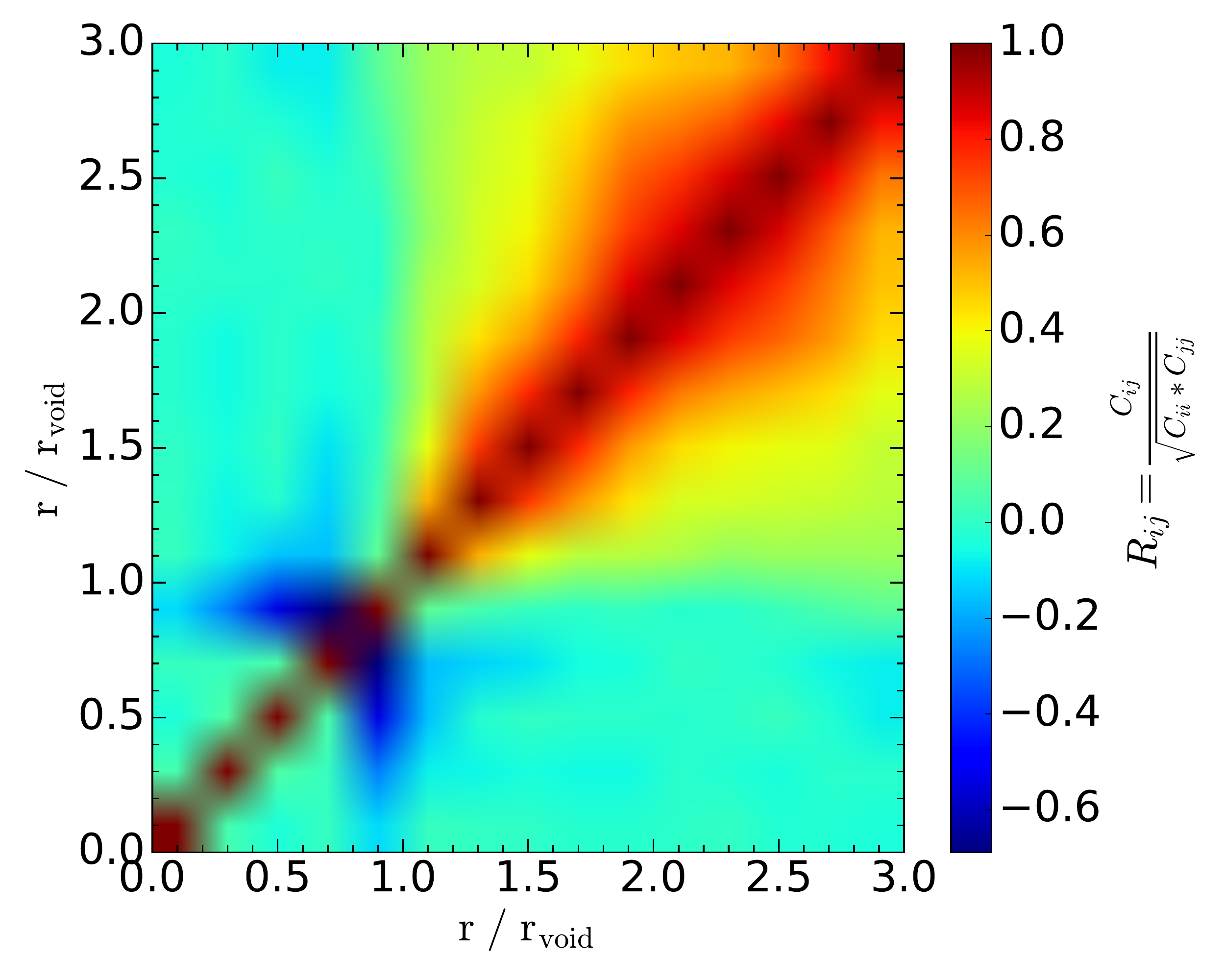}
    \caption{The correlation matrix of the subhalo density profile measured for voids traced by total-mass
     selected subhaloes in Ref-L0100N1504 (black solid line in Fig. \ref{Figure: Density profiles}.
     The colourmap shows the Pearson product-moment correlation coefficients for each bin
     in the density profile. There are not many strong off-diagonal correlations among different bins,
     which means that we can compare the profiles in Section \ref{Section: Void profiles}
     with the error bars shown shown in there to get an estimate
     of the level at which these the differences are statistically significant.}
    \label{Figure: Covariance matrix}
    \centering
\end{figure}

\begin{figure}
	\includegraphics[width=\columnwidth]{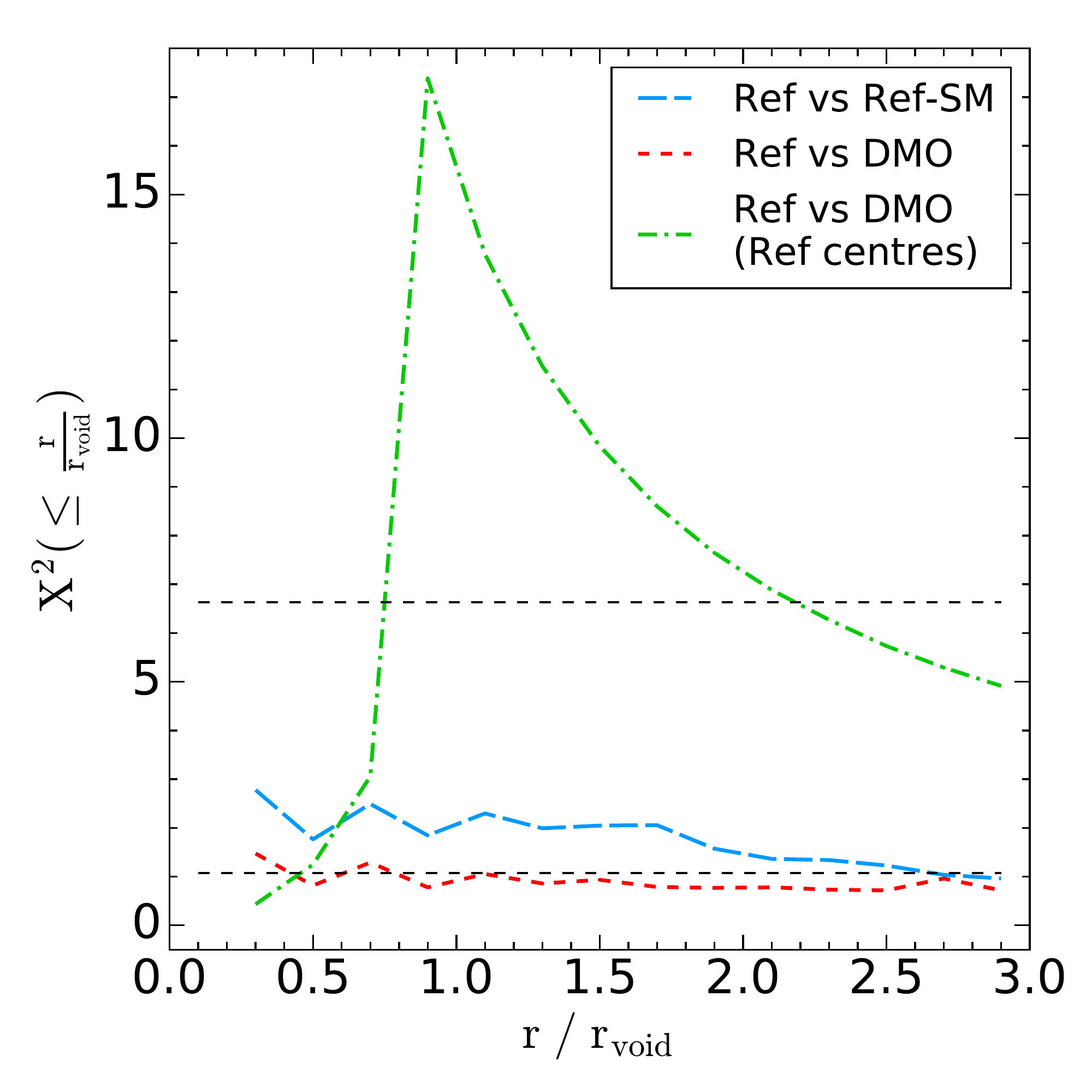}
    \caption{The value of the reduced $\chi ^2$ statistic (\ref{Equation: chi2}) comparing
    the density profiles of Fig. \ref{Figure: Density profiles}, as obtained when considering
    only the values of the profiles up to $\rm{r\ /\ r_{void}}$. The different lines
    compare the density profile inferred from subhaloes selected by their total mass
    with the profile obtained from: blue, long-dashed line: subhaloes selected by their
    stellar mass in Ref-L0100N1504. Red dashed line: Subhaloes selected by their total
    mass in DM-L0100N1504. Green, dot-dashed line: Subhaloes selected by their total
    mass in Ref-L0100N1504, but having their profiles measured in DM-L0100N1504.
    The null hypothesis is that each pair of profiles come from the same distribution. 
    The top and bottom constant, dashed lines show values at which the null hypothesis
    can be rejected with a 90 and 70 per cent level of confidence, respectively.}
    \label{Figure: chi2}
    \centering
\end{figure}

In Sec. \ref{Section: Void profiles} we showed measurements of the radial density and velocity profile for voids identified using subhaloes in the EAGLE simulations. The error bars in Fig. \ref{Figure: Density profiles} and Fig. \ref{Figure: Velocity profiles} show the scatter around the mean density profiles, obtained using a jackknife re-sampling of all the voids used in the stacked (averaged) profiles. Due to the sensitivity of the profiles to the number and the location of the bins, there could be correlated errors among values from different bins, which could result in an under-estimation of the true errors of the profiles. Thus, even though two profiles may show differences that are bigger than the error bars, these differences could still not be statistically significant when the correlated errors are taken into account. To explore this in further detail we compute the correlation matrices of all the curves shown in Fig. \ref{Figure: Density profiles} and Fig. \ref{Figure: Velocity profiles}. In Fig. \ref{Figure: Covariance matrix} we show the correlation matrix for the subhalo density profile of voids traced by subhaloes selected by their total mass (black, solid line in the left panel of Fig. \ref{Figure: Density profiles}). Correlation matrices of the remaining profiles are very similar and are not shown here.
It appears that there are no strong off-diagonal correlations among bins, which means that it is safe to compare the differences in void profiles with the error bars shown in there.

We can address the problem of whether the differences in the profiles of Fig. \ref{Figure: Density profiles} arise just because of chance by computing the reduced $X^2$ statistic, generalized for data with correlated errors:

\begin{align} \label{Equation: chi2}
X ^2 = \sum_{i,j} \rm{C}_{ij}^{-1} \frac{\left( x_{i_1} - x_{i_2})(x_{j_1} - x_{j_2} \right)}{\rm{dof}}\ \ ,
\end{align}

\noindent where $\rm{C}^{-1}$ corresponds to the inverse matrix of the sum of the covariance matrices of the two curves that are being compared, $x_{ik}$ is the mean value of the density profile of curve $k$ in the bin $i$, and dof stands for degrees of freedom. Under the null hypothesis that two profiles come from the same statistical distribution, we can measure of how probable it is to obtain differences as seen in  Fig. \ref{Figure: Density profiles} just because of statistical fluctuations, by computing the $X^2$ statistic.

Fig. \ref{Figure: chi2} shows the value of $X^2$ when comparing different subhalo density profiles (Fig. \ref{Figure: Density profiles}), but only using the values of the profile up to $r\ /\ r_{\rm{void}}$. Thus, the last value at the right-hand end of the curves show the obtained $X^2$ using all the bins in the calculation. The different lines compare the density profile of voids traced by subhaloes selected by their total mass in Ref-L0100N1504. The blue long-dashed line shows a comparison with voids traced by stellar mass-selected subhaloes, while the red dashed and the green dot-dashed lines show comparisons with voids traced by total mass-selected subhaloes in DM-L0100N1504, and voids found in Ref-L0100N1504 but with their profiles measured in Ref-L0100N1504. The horizontal dashed lines shows values of $X^2$ where the null hypothesis can be rejected with a 70 (bottom) and 90 (top) per cent confidence level. The differences between the profiles of voids traced by total mass-selected subhaloes and those traced by stellar mass-selected subhaloes in Ref-L0100N1504 are statistically significant with a 70 per cent confidence level (blue, long-dashed line). The same holds true for a comparison between voids in Ref-L0100N1504 and DM-L0100N1504 when the same total mass selection cut is used for the tracers (red dashed line). When the same void centres are used to compute the density profile in Ref-L0100N1504 and DM-L0100N1504, very small differences are detected, but these are statistically significant to a 75 per cent confidence level.


\bsp	
\label{lastpage}
\end{document}